\documentclass[a4paper,fleqn]{cas-dc}

\usepackage[numbers]{natbib}
\usepackage{subfig}
\usepackage{tabularx}
\usepackage{subfloat}
\usepackage{multicol}
\usepackage{supertabular}
\usepackage{amsmath,amssymb,amsfonts}
\usepackage{algorithmic}
\usepackage{graphicx}
\usepackage{textcomp}

\def\tsc#1{\csdef{#1}{\textsc{\lowercase{#1}}\xspace}}
\tsc{WGM}
\tsc{QE}
\tsc{EP}
\tsc{PMS}
\tsc{BEC}
\tsc{DE}
\begin{document}
\let\WriteBookmarks\relax
\def\floatpagepagefraction{1}
\def\textpagefraction{.001}
\shorttitle{Augmented Democracy in Smart Cities}
\shortauthors{E. Pournaras}

\title[mode = title]{Proof of Witness Presence: Blockchain Consensus for Augmented Democracy in Smart Cities}                      
%\tnotemark[1,2]

%School of Computing, University of Leeds, Leeds LS2 9JT, UK, Tel.: +441133435447, \email{e.pournaras@leeds.ac.uk}

%\tnotetext[1]{This document is the results of the research
%   project funded by the National Science Foundation.}

%\tnotetext[2]{The second title footnote which is a longer text matter
%   to fill through the whole text width and overflow into
%   another line in the footnotes area of the first page.}

\author{Evangelos Pournaras}[]
\cormark[1]
\ead{e.pournaras@leeds.ac.uk}
%\ead[url]{www.cvr.cc, cvr@sayahna.org}

%\credit{Conceptualization of this study, Methodology, Software}

\address[1]{School of Computing, University of Leeds, Leeds LS2 9JT, UK}

\cortext[cor1]{Corresponding author}

\begin{abstract}
Smart Cities evolve into complex and pervasive urban environments with a citizens' mandate to meet sustainable development goals. Repositioning democratic values of citizens' choices in these complex ecosystems has turned out to be imperative in an era of social media filter bubbles, fake news and opportunities for manipulating electoral results with such means. This paper introduces a new paradigm of augmented democracy that promises actively engaging citizens in a more informed decision-making augmented into public urban space. The proposed concept is inspired by a digital revive of the Ancient Agora of Athens, an arena of public discourse, a Polis where citizens assemble to actively deliberate and collectively decide about public matters. The core contribution of the proposed paradigm is the concept of proving witness presence:  making decision-making subject of providing secure evidence and testifying for choices made in the physical space. This paper shows how the challenge of proving witness presence can be tackled with blockchain consensus to empower citizens' trust and overcome security vulnerabilities of GPS localization. Moreover, a novel platform for collective decision-making and crowd-sensing in urban space is introduced: Smart Agora. It is shown how real-time collective measurements over citizens' choices can be made in a fully decentralized and privacy-preserving way. Witness presence is tested by deploying a decentralized system for crowd-sensing the sustainable use of transport means. Furthermore, witness presence of cycling risk is validated using official accident data from public authorities, which are compared against wisdom of the crowd. The paramount role of dynamic consensus, self-governance and ethically aligned artificial intelligence in the augmented democracy paradigm is outlined. 
%This template helps you to create a properly formatted \LaTeX\ manuscript.
%
%\noindent\texttt{\textbackslash begin{abstract}} \dots 
%\texttt{\textbackslash end{abstract}} and
%\verb+\begin{keyword}+ \verb+...+ \verb+\end{keyword}+ 
%which
%contain the abstract and keywords respectively. 
%
%\noindent Each keyword shall be separated by a \verb+\sep+ command.
\end{abstract}

%\includegraphics{figs/grabs.pdf}
%\item An augmented democracy paradigm for Smart Cities using blockchain consensus
%\item Collective decision-making subject of proving witness presence in urban space
%\item Proving witness presence: secure verification of location, time, situation awareness
%\item Review of participatory systems and blockchain-based localization systems
%\item Decentralized and privacy-preserving collective measurements maps
%\item Proof of concept: sustainability evaluation of mobility and validation of cycling risk


\maketitle

\section{Introduction}\label{sec:introduction}

Smart City urban environments co-evolve to complex informational ecosystems in which citizens' collective decisions have a tremendous impact on sustainable development. Choices about which transport mean to use to decrease noise levels or carbon emissions, which urban areas may require gentrification or new policies for improving safety are some examples in which decision-making turns out to be complex and dynamic~\cite{Salesses2013}. It is apparent that the 4-year electoral agendas of political parties based on which they unfold their policies are either impractical or outdated for such urban ecosystems. Policy-making, participation and ultimately democracy requires a revisit and a digital transformation for the better of citizens. 

Existing social media platforms, powered by citizens' personal data and centralized machine learning algorithms can isolate citizens via informational filters bubbles and manipulate them using fake information~\cite{Vosoughi2018,Lazer2018}. Citizens often feel powerless to influence public matters and, beyond elections, there is no established channel for their voice to be heard in centers of decision-making~\cite{Fung2015}. Despite the technological capabilities to engage wisdom of the crowd for decision-making, decisions remain to a high extent top-down and political actions do not always align with electoral political agendas~\cite{Gibson2016}. The rise of populism, extremism and electoral manipulations showcase the risks of democratic values in decay~\cite{Emerson2020}.

To address these challenges a new digital paradigm of augmented democracy is introduced to empower a more informed, engaging and responsible decision-making au\-gme\-nted into public urban space, where the decisions have a direct impact. In this sense, augmented democracy is envisioned as a digital revive of the Ancient Agora of Athens, a public assembly of citizens for discourse, deliberation and collective decisions-making. Witness presence has been so far the missing but required value in digital democratic processes: the act of intervening and testifying about the physical world as well as the undertaking of responsibility for these actions. For instance, making the rating of traffic congestion at different streets conditional to secure digital evidence about the citizen's location and speed records at these streets is an example of proving witness presence. Validating such digital evidences without relying to a trusted third party is a highly inter-disciplinary and complex challenge involving research from the areas of distributed systems, security, Internet of Things, social science, mechanism design and others~\cite{Amoretti2018,Qi2017,Jiang2020,Falk2018,Allen2017}.

The envisioned scenario is the following: Citizens navigate over several urban points of interest with augmented information. They make more informed and trustworthy choices by proving witness presence in one of these points. They also access live updates about the collective choices made by other citizens in relevant points of interests. This paper shows how this challenging scenario can be made technically feasible and viable using secure, privacy-preserving and decentralized information systems, e.g. blockchain consensus, as well as crypto-economic design principles to incentivize participation, engagement, while limiting adversary behavior. The proposed solution consists of the three following pillars: (i) \emph{participatory crowd-sensing}, (ii) \emph{proof of witness presence} and (iii) \emph{real-time collective measurements}. Despite the complexity and ambition level of the proposed endeavor, this paper demonstrates a first prototyped system (testnet) that integrates and deploys all three pillars. It also illustrates a use case scenario on cycling safety that validates the quality of information acquired via citizens' witness presence using official data from public authorities. The role that dynamic consensus, self-governance and artificial intelligence play in the proposed augmented democracy paradigm is discussed. 

Compared to related initiatives such as online pe\-ti\-tion/vo\-ting systems~\cite{Bungale2003,Perg2017}, promising participatory budgeting initiatives for more equitable and transparent distribution of resources~\cite{World2008} as well as other e-participation approaches~\cite{Zolotov2018}, the proposed augmented democracy paradigm fundamentally differs in the following aspects: (i) It does not rely on trusted third parties. (ii) It can operate in real-time and is not limited to long-term decision-making.(iii) It encourages a more informed and responsible decision-making by better integrating citizens' choices into daily life and public space. (iv) It has a broader inter-disciplinary scope and applicability. In summary, the contributions of this paper are outlined as follows: 

\begin{itemize}
\item A new three-tier paradigm of augmented democracy in Smart Cities.
\item The Smart Agora crowd-sensing platform for modeling complex spatio-temporal crowd-sensing scenarios of augmented decision-making. 
\item The new blockchain consensus concept `proof of witness presence' and a study of how it is technically realized.
\item A review of related initiatives on digital democracy as well as blockchain-based approaches for proof of location. 
\item The concept and realization of `collective measurements maps' that filter out geolocated data and determine the points of interest from which data are aggregated.
\item A first fully-fleshed working prototype of the augmented democracy paradigm meeting minimal requirements set for a proof of concept. 
\item A use case scenario on cycling safety demonstrating the capacity of citizens' witness presence to match accurate information from official public authorities. 
\end{itemize}

This paper is outlined as follows: Section~\ref{sec:related-work} outlines the theory and current practice behind digital democracy initiatives. Section~\ref{sec:paradigm} introduces the vision and challenges of the augmented democracy paradigm that consists of three pillars. The first pillar of participatory crowd-sensing is illustrated in Section~\ref{sec:crowd-sensing}. The concept of proving witness presence is introduced in Section~\ref{sec:consensus} that is the second pillar of the proposed paradigm. The third pillar of real-time collective measurements is introduced in Section~\ref{sec:collective-measurements}. The evaluation methodology and experimental results are illustrated in Section~\ref{sec:evaluation}. Section~\ref{sec:discussion} discusses dynamic consensus and self-governance as well as the role of artificial intelligence for augmented democracy. Finally, Section~\ref{sec:conclusion} concludes this paper and outlines future work.

%The viability of direct democracy has been often criticized for inefficient or unstable outcomes as well as vulnerabilities related to voting manipulation, tyranny of the majority and others.  

%Particularity of proof of location compared to other consensus mechanisms: the proximity of the validators matters whereas in others does not, they only require a communication channel
%
%
%\cite{Chen2017}
%
%Survey
%
%
%%%%%%%%%%%%%%%%%%%%%%%%%%%%%%%%%%%%%%%%%%%%%%%%%
%
%\cite{Zidek2018}
%
%Bellrock for security and privacy a solution based on anonymous beacons.
%
%
%%%%%%%%%%%%%%%%%%%%%%%%%%%%%%%%%%%%%%%%%%%%%%%%%
%
%\cite{Nissen2018}
%
%Geocoin
%
%%%%%%%%%%%%%%%%%%%%%%%%%%%%%%%%%%%%%%%%%%%%%%%%%
%
%\cite{Nasrulin2018}
%
%Proof of location protocol
%
%%%%%%%%%%%%%%%%%%%%%%%%%%%%%%%%%%%%%%%%%%%%%%%%%
%
%\cite{Dasu2018}
%
%Unchain your blockchain
%
%
%%%%%%%%%%%%%%%%%%%%%%%%%%%%%%%%%%%%%%%%%%%%%%%%%

\section{Theoretical Underpinning and Related Work}\label{sec:related-work}

Political philosophers and democratic theorists have argued that delegating the `right of sovereignty' could not be democratic resulting in aristocracy as well as in non-political and illegitimate state~\cite{Urbinati2008}. The proposed augmented democracy approach suggests new pathways to diminish this delegation, and reclaim sovereignty at a local and community level. The higher feasibility of a `renewed version of democratic representation' based on `smaller, decentralized, and distributed (offline and online) citizen assemblies' is earlier hypothesized as the means to guarantee legitimacy when rea\-ching mass participation is challenging~\cite{Poblet2017,Aitamurto2014}. A more localized scope in collective decision-making can also mitigate the trilemma of democratic reform~\cite{Fishkin2011}: among the principles of \emph{political equality}, \emph{mass participation} and \emph{deliberation}, promoting any of the two, hinders the third. In particular, the current online crowd-civic platforms can only address highly engaged deliberators. As such they cannot represent well the broader population and, in this sense, guarantee political equality. 

Earlier contemporary theory has also suggested that while represented democracy is technically feasible, it remains an oxymoron, in contrast to direct democracy that comes as the norm but impractical~\cite{Urbinati2004}. A proposed horizontal and acephalous political order suggests legislative power held by multiple actors and functioning within elected and citizen assemblies at multiple times and spaces. Citizens come with both electoral rights and rights to revoke or censure laws~\cite{Condorcet1793}. This approach aspires to reconcile sovereignty, representation, and participation with the latter settling a `sou\-rce of stability and innovation', while representation is the means to collect data and knowledge for public interest~\cite{Ober2008,Urbinati2004}. New opportunities arise to experimentally test novel radical ideas that have been so far approached by researchers on a more theoretical basis, for instance, quadratic voting~\cite{Lalley2018,DemocracyEarth2020} or a more egalitarian ranking aggregation of voting solutions~\cite{Contucci2016,Emerson2020}. 

Most research efforts on digital democracy focus on online petitions, voting and the design of collaboration platforms for deliberation and collective decision-making. For instance, WeCollect~\cite{WeCollect2019} is a Swiss independent non-profit platform that moderates networking of citizens, collects signatures for popular initiatives and referendums including topics such as refugees, basic income, energy policies and other. Such efforts are also observed within the Zurich Political Participation~\cite{StadtZurich2019} portal that administrates online petitions and self-initiatives published in newspapers. Such efforts based on online petitions fundamentally differ from the proposed augmented democracy paradigm as they are not designed for real-time feedback and interactions. Instead they aim to increase participation into existing established democratic processes and provide new representation means to various social groups. 

CONSUL~\cite{CONSUL2019,Pena2017} is an open-source citizens' participation software that supports open, transparent and democratic governance. The software supports debates, citizen proposals, participatory budgeting, voting and collaborative legislation. CONSUL has been extensively used by city authorities and organizations all over the world with several local projects featured online~\cite{CONSUL2019}. Further progress of such democratic initiatives in Spain has resulted in the open-source participation solution of Decidim~\cite{Decidim2019,Aragon2017} that configures participation spaces such as initiatives, assemblies, processes and consultations supported by face-to-face meetings, surveys, proposals, voting and other. More specifically, the assembly spaces provide the option of geolocating periodic meetings, whose composition and agenda are self-organized by participants. These two state-of-the-art platforms as well as DemocracyOS~\cite{DemocracyOS2019} could benefit and work in synergy with the proposed augmented democracy solution as it can position more effectively collective decision-making in citizens' daily life and the public space they experience. 

There are other platforms with a narrower scope and focus. For instance, Crossiety~\cite{Crossiety2019} is a startup with a mobile app implementing social networking functionality to connect local communities and villages. Airesis~\cite{Airesis2019} is an online deliberation tool that manages citizens' shared proposals and debates. It supports temporary anonymity, secret ballot, auditable voting and the Schulze voting method~\cite{Schulze2011}. Deliberatorium~\cite{Deliberatorium2019} is designed to support crowds to deliberate and have productive discussions about complex problems. It combines argumentation theory and social computing in a web-based system to promote dialogue, citizens' retention and engagement~\cite{Iandoli2018}. In contrast to the aforementioned deliberation and other engagement platforms~\cite{Novoville2019,Discourse2019,Adhocracy2019}, the augmented democracy approach of this paper moves a step forward by addressing quality aspects on collective decision-making by empowering proof of claims and testimonies in citizens' choices. 

Crowd-sensing and citizen science initiatives can also provide insights and empirical evidence to policy makers. For instance, Place Pulse~\cite{PlacePulse2019,Dubey2016,Naik2014,Salesses2013} is a platform for mapping and measuring quantitatively urban qualities in cities as perceived by citizens. Such qualities include how wealthy, modern, safe, lively, active, unique, central, adaptable or family friendly an urban space is. Another environmental initiative is CrowdWater~\cite{CrowdWater2019,Seibert2019} that is designed to collect data about the water level, soil moisture and tempora\-ry stre\-ams to predict floods and water flows. None of the above initiatives is designed for direct online decision-making, nevertheless, the domain data they harvest can be used as empirical evidence in the proposed augmented democracy paradigm. 

Finally, blockchain solutions for participatory and democratic processes are subject of active research~\cite{Qi2017,Susskind2017,DemocracyEarth2020}. A\-go\-ra~\cite{Agora2019} and Follow My Vote~\cite{FollowMyVote2019} rely on a decentralized voting protocol and consensus mechanism to establish secure and transparent ballots as well as voting results that are publicly verifiable. Democracy Earth~\cite{DemocracyEarth2020} focuses on a censorship-resistant social layer on top of distributed ledgers. It runs intersubjective consensus~\cite{Wan2010} that uses social markers to incentivize participation on the blockchain economy and earn rights. The system is designed to deploy borderless democracies, universal basic income mechanisms and credit scores, without the need to sacrifice privacy. Votetandem~\cite{Votetandem2019} is based on blockchain technology with which Swiss citizens can supply their vote to inhabitants in Switzerland excluded from voting, e.g. foreigners making up 25\% of the population. However such voting solutions have not yet integrated in the public urban space and do not focus on a higher situation awareness in collective decision-making. 

\section{Augmented Democracy: Vision and Challenges}\label{sec:paradigm}

This paper envisions a digital revive of the ancient \emph{agora} of Athens, a public cyber-physical arena of discourse, where citizens actively assemble, deliberate and engage in informed collective decision-making about a wide range of complex public matters. The scenario envisioned is the following: Individual citizens, regional communities or policy makers crowd-source complex decision-making processes augmented into Smart Cities, for instance, decide how to better integrate immigrants, how to improve public safety or transport means, how to deal with gentrification and others. Such processes are designed to encourage or even enforce a more informed and participatory decision-making to improve indi\-vi\-dual\-/\-co\-lle\-cti\-ve awareness and the quality of decision outcomes. In practice this means that a citizen with a community mandate to participate in a collective decision-making process uses a smart phone and navigates in the urban environment to visit or discover \emph{points of interests} with augmented information. For instance, after a natural disaster, i.e. flooding, earthquake, etc., citizens can rate the severity of damages at different locations to orchestrate mitigation actions more effectively. Citizens have a saying, an informed one, backed up by evidence of \emph{witness presence} in the cyber-physical space of Smart Cities. Witness presence is an added value on citizens' decision-making created at a \emph{certain location}, at \emph{certain time} with a \emph{certain situation awareness} when performing a \emph{certain action}. Such evidence-based collective decision-making process introduces highly contextuali\-zed spa\-tio-te\-mpo\-ral data, whose aggregation creates a live pulse of the city, a public good created by citizens, for citizens. For instance, live updates about the severity of damages in certain areas can engage remote volunteers for support or act as warning signals for civilians to avoid these areas and protect their life.  

Such a scenario of a direct augmented democracy in Smart Cities requires data-intensive information systems playing a key role for the viability and trust of this challenging endeavor. A centralized design for these critical systems can pose several undermining risks: (i) Existing centrally managed online social media, along with traditional media, are often carriers of unaccountable and uncredible information that is a result of manipulative nudging and spreading of fake news~\cite{Vosoughi2018,Lazer2018}. The damage in the participation level and trust of citizens on democratic processes, such as elections and referendums, can be unprecedented~\cite{Leduc2015,Grvcar2017}. (ii) The most prominent global localization service, the GPS, is centrally controlled, it has several security and privacy vulnerabilities, i.e. spoofing and jamming~\cite{Tippenhauer2011}, it is not accurate enough and has restricted coverage, e.g. indoor localization is not feasible~\cite{Nissen2018}. (iii) Collective measurements and awareness via Big Data analytics rely on trusted third parties that are single point of failure. They usually collect and store personal sensitive data and as a result profiling and discriminatory actions over citizens become feasible.

This paper claims that in principle any digital democracy paradigm cannot remain viable in the long term unless the management of information systems is democratized. As democracies cannot properly function even with benevolent totalitarian forces, similarly, centralized information systems for governance, however well they perform and simple to manage, they can always be subject of manipulation and misuse in such a critical service for society.

\begin{figure}
\centering	
\includegraphics[width=1.0\columnwidth]{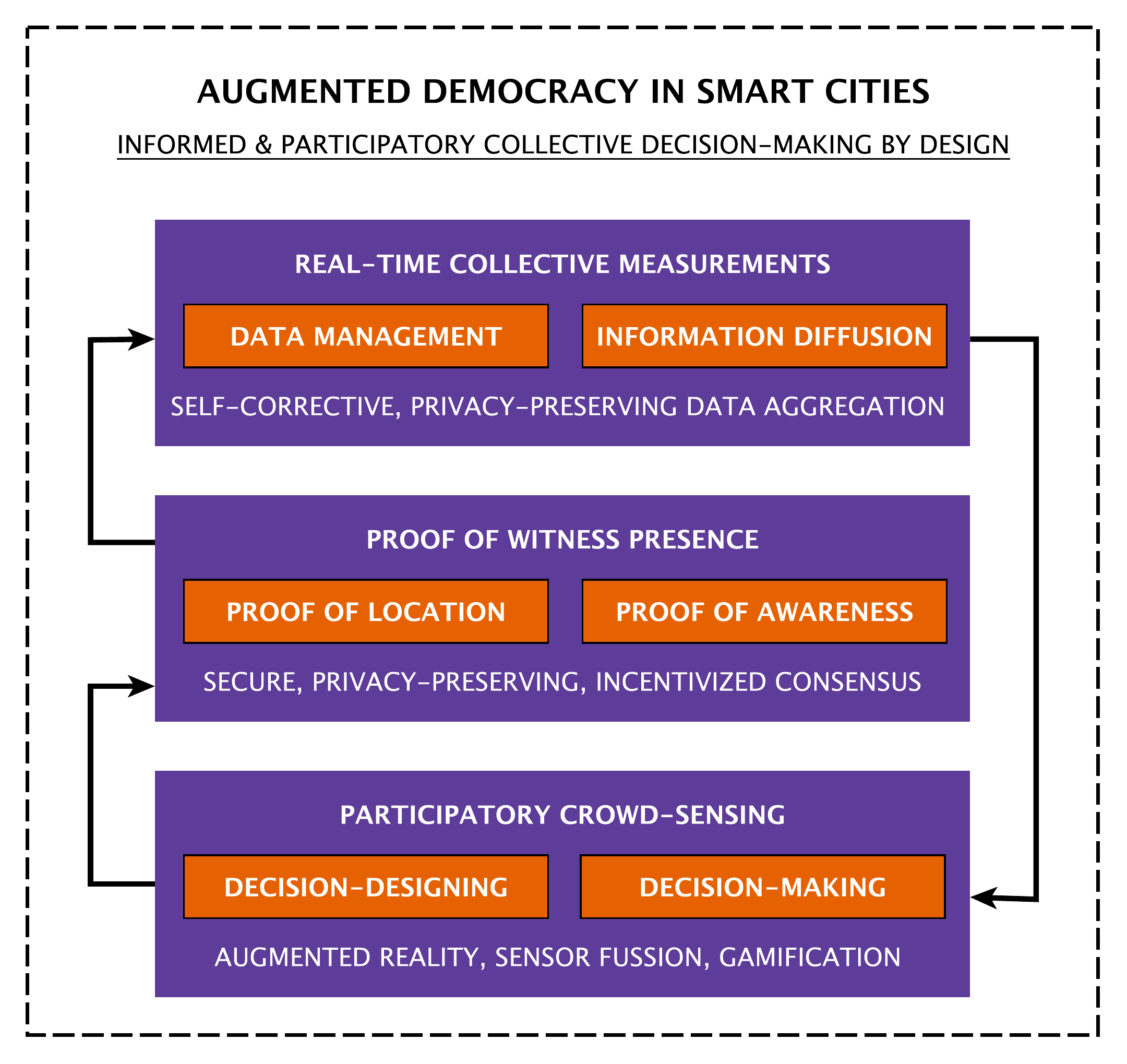}
\caption{An augmented democracy paradigm for Smart Cities consisting of three pillars: (i) Crowd-sensing is performed within participatory witness presence scenarios of augmented reality in public spaces. (ii) Proof of witness presence is performed by securely verifying the location and the situation awareness of citizens without revealing privacy-sensitive information. (iii) Real-time and privacy-preserving collective measurements are performed, subject of witness presence.}\label{fig:architecture}
\end{figure}

The positioning of this paper is that decentralized information systems, particularly distributed ledgers, consensus mechanisms and crypto-economic models, can by used to design a more informed and participatory collective decision-making as shown within the three pillars of Figure~\ref{fig:architecture}. This is possible by introducing the concept of \emph{witness presence} as a consensus model for verifying location and situation awareness of collective decision-making in Smart Cities.

Each pillar involves a technical challenge addressed in this paper: (i) How to design a general-purpose crowd-sensing system for the Internet of Things to reason about the quality of decision-making in public space. (ii) How collective decision-making can be made conditional of proving witness presence using blockchain consensus to empower trust. (iii) How to access real-time spatio-temporal collective measurements made in decentralized and privacy-preserving way as a result of witness presence. The rest of this paper illustrates each of the three pillars in the proposed framework of augmented democracy.

\section{Participatory Crowd-sensing}\label{sec:crowd-sensing}

\begin{figure}
\centering	
\includegraphics[width=1.0\columnwidth]{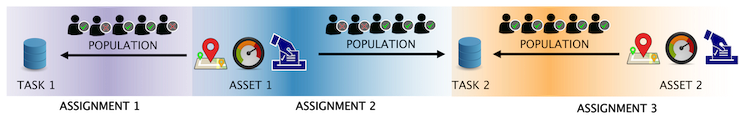}
\caption{Modeling a crowd-sensing project with Smart Agora. A \emph{project} consists of one or more \emph{assets}, \emph{tasks}, and \emph{assignments}. (i) An asset defines complex crowd-sensing processes and consists of configurations about the point of interests, the questions and the collected sensor data. (ii) A task stores and manages the collected citizens' data as defined by an asset. (iii) An assignment links together an asset and a task and launches the crowd-sensing process by selecting candidate citizens for participation. In this visual example, Task 1 results in crowd-sensing data from the Assignment 1 of Asset 1 to a sample of the population. In contrast, Task 2 is the result of Assignment 2 of Asset 1 to a different population sample as well as Assignment 3 of Asset 2 to the whole population.}\label{fig:Smart-Agora-project}
\end{figure}

At the foundations of the framework lies the award-win\-ning\footnote{Smart Agora has been part of the Empower Polis project that won the 1st prize at the ETH Policy Challenge~\cite{ETHPolicyChallenge2019}.} platform of Smart Agora, a pillar that empowers citizens to (i) visually design and crowd-source complex decision-making processes augmented in the urban environment as well as (ii) make more informed decisions by witnessing the urban environment for which decisions are made. Figure~\ref{fig:Smart-Agora-project} outlines how an augmented democracy project is modeled\footnote{The modeled entities follow the concept of Hive~\cite{Hive2019}.}. 

\begin{figure}
\centering	
\subfloat[Determining augmented point of interests with survey questions.]{\includegraphics[width=1.0\columnwidth]{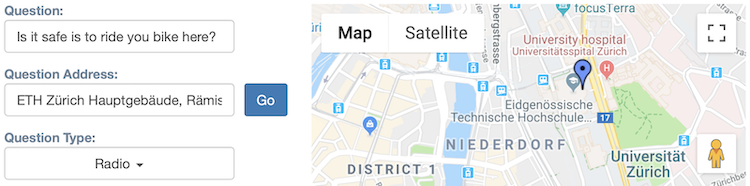}}\\
\subfloat[The Smart Agora App]{\includegraphics[width=1.0\columnwidth]{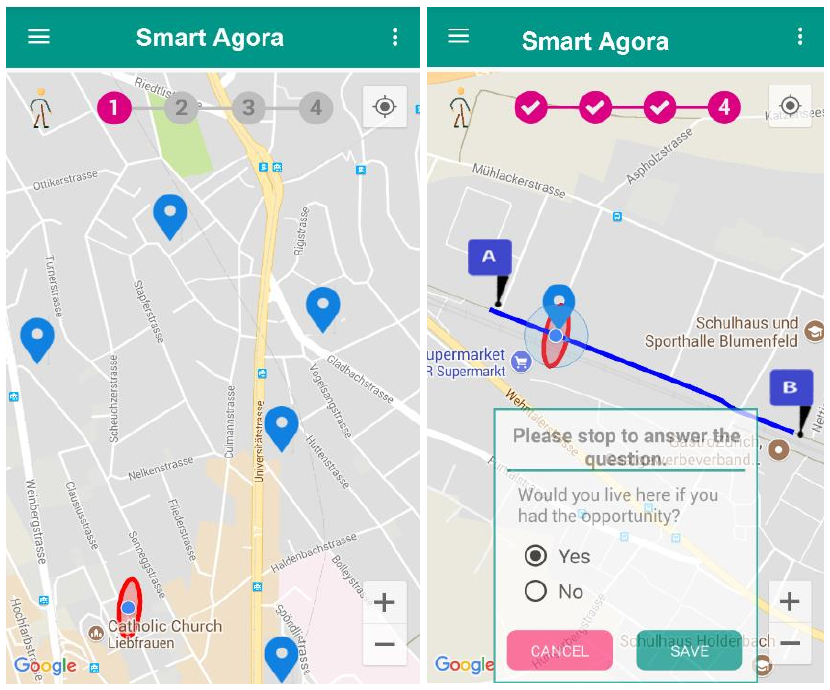}}
\caption{The Smart Agora software platform.}\label{fig:Smart-Agora-software}
\end{figure}

Decision-making processes are designed in a visual and interactive way as follows: A number of \emph{points of interest} are determined in an interactive map as shown in Figure~\ref{fig:Smart-Agora-software}a. Each point of interest hosts a number of questions\footnote{Radio, checkbox, likert and text box questions are currently supported.} that citizens can answer on their smart phone if and only if they are localized nearby the point of interest (see Figure~\ref{fig:Smart-Agora-software}b). An ellipse~\cite{Griego2017} with configurable size is determined around each moving citizen. Localization is performed when a point of interest falls in the ellipse, triggering an event that prompts citizens to answer questions on their smart phone based on what they witness in the public urban space they are located that moment. 

Points of interest can be given by an oracle~\cite{Lo2020,Ma2019}, i.e. a policy maker running a specific voting campaign, or they can be crowd-sourced to communities based on crypro-economic incentive models. For instance, FOAM~\cite{FOAM2019} relies on token curated registries~\cite{Ramachandran2018,Falk2018} that realize economic and reputation incentives for citizens to play the role of cartographers and contextualize crypto-spatial coordinates\footnote{On-chain and off-chain verifiable location information of FOAM consisting of a geohash and an Ethereum smart contract address. It can approximate resolution of one square meter that allows a maximum of 500 trillion unique addresses.} with meta-info\-rma\-tion.

Each question as well as their possible answers can be incentivized with rewards in the form of different crypto-currencies, i.e. utility tokens used for a value exchange required to run and incentivize the augmented democracy pa\-ra\-digm. For instance, tokens created by a city council to incentivize participation in a crowd-sensing project for improving the quality of public transport can be collected and used by citizens to purchase public transport tickets. Similarly parking away from crowded city centers can be incentivized with tokens that can issue discounts in nearby shops. Sensor data can also be periodically collected and used for supporting the two above pillars in Figure~\ref{fig:architecture}, i.e. sensor fusion to prove claims of witness presence~\cite{Wolberger2018} or aggregation measurements over sensor data can be performed to increase collective awareness~\cite{Pournaras2017c}. 

A decision-making process can be designed in three \emph{navigation modalities}: (i) \emph{Arbitrary}--the points of interests can be arbitrary visited by citizens. Questions are always triggered whenever citizens visit a new point of interest. (ii) \emph{Sequential}--A sequence is determined for visiting the points of interests. Only the questions of the next point of interest can be triggered, imposing in this way an order. (iii) \emph{Interactive}-- The next point of interest is determined by the answer of the citizen in the current point of interest. The latter modality can serve more complex decision-making processes as well as gamification scenarios.

\section{Proof of Witness Presence}\label{sec:consensus}

Witness presence provides an added value in participatory decision-making~\cite{Lam2015,Nevejan2009}. Witnessing public happenings and the complex urban environment of Smart Cities empowers a Polis of active citizens that can directly influence real-world by intervening and testifying instead of remaining passive spectators of a reality for which others decide, a limitation of current representative democracies. Ultimately, witness presence is about encouraging the taking of responsibility on spot, a requirement for a viable democracy. While witness presence can be seen as a political statement, it is actually a highly complex techno-socio-economic problem in the context of the proposed augmented democracy paradigm: \emph{Proving of being present at a certain location, at a certain time with a certain situation awareness in order to perform certain actions, while having the incentive to participate}. Section~\ref{subsec:proof-of-location} and~\ref{subsec:witness-presence} review blockchain consensus models for location proofs and social proofs respectively. Section~\ref{subsec:blockchain} also illustrates their synthesis into a blockchain consensus network for proving witness presence.

\subsection{A review on proof of location}\label{subsec:proof-of-location}

At the core of witness presence lies \emph{proof of location} that is the secure verification of a citizen's spatial position. It requires accurate estimation of distances or angles of signals exchanged between {wireless transmitting devices. These distances are calculated by measuring signal attenuation or signal propagation times. Techniques of the former, i.e. Received Signal Strength Indicator (RSSI)~\cite{Musciotto2016}, are common but do not provide accurate estimates, while techniques of the latter, i.e. Time of Flight (ToF) with algorithms based on triangulation, trilateration or multilateration, require synchronized clocks to eliminate clock drifts of the oscillators~\cite{Chen2017}. For example, the Global Positioning System (GPS) relies on high-precision atomic clocks on satellites that synchronize with centralized master control stations on the ground. Recently, decentralized algorithms for Byzantine fault-tolerant clock synchronization have been studied~\cite{Malekpour2015,Khanchandani2019}. These algorithms run by autonomous interactive wireless receivers and transmitters, i.e. beacons, that self-determine via their communication the geometry of their zone coverage without third parties. By reaching an agreement about a common time\footnote{Not necessarily a UTC time unless some oracle information is used.}, specific locations can be accurately detected via trilateration~\cite{Malekpour2017}. 

The proof of location required for the proof of witness presence can be achieved with various trade-offs using one or more of the following infrastructures: (i) \emph{GPS}, (ii) \emph{mobile cellular network}, (iii) \emph{low power wide area network} (LPWAN) and (iv) \emph{peer-to-peer ad hoc (opportunistic) networks} consisting of several different Internet of Things devices such as smart phones, static beacons, wearables, wireless access points, etc. Table~\ref{table:proof-of-location} summarizes a comparison of the block\-chain-ba\-sed approaches for proof of location.

\begin{table*}
\caption{A comparison of blockchain-based approaches for proof of location based on criteria that can make the augmented democracy paradigm more viable.}\label{table:proof-of-location}
\centering
\resizebox{\textwidth}{!}{%
\begin{tabular}{r l l l l}%{l l l l l}
\toprule
Approaches & GPS~\cite{Nissen2018} & Mobile Cellular Network~\cite{Victor2018} & LPWAN~\cite{FOAM2019} & P2P Ad Hoc Networks~\cite{Amoretti2018} \\\addlinespace\toprule
Infrastructure-independent & No & No	& No & Yes \\\midrule
Decentralization & Low &	Low	& Medium & High \\\midrule
Access & Open & Closed & Open & Open \\\midrule
Management & Governmental-level & Enterprise-level & Community-level & Self-organized \\\midrule
Disaster Resilience & Medium & Medium & Medium & High \\\midrule
Coverage Range & Global & National & Urban & Localized \\\midrule
Indoor Coverage & No & Yes & Yes & Yes \\\bottomrule
\end{tabular}
}
\end{table*}

On the one hand, GPS is a free service with planetary coverage and as such it can be easily used by a Smart Agora application for outdoor localization, as the current prototype supports. Similarly, GeoCoin relies on GPS for the location-based execution of smart contracts~\cite{Nissen2018}. However, GPS is a single point of failure, it is highly susceptible to fraud, spoofing, jamming and cyber-attacks, it does not provide any proof of origin or authentication and therefore it is unreliable by itself to prove claims of locations. Moreover, GPS cannot provide indoor localization, it underperforms in high density urban environments, i.e. increased signal multipath, and its energy consumption is prohibitive for low-power devices. Such vulnerabilities have been prominently identified in smart watches\footnote{Such vulnerabilities have been demonstrated by a German security researcher after a smart watch vendor ignored vulnerability reports for more than a year, leaving thousands of GPS-tracking watches open to attackers~\cite{PWNEDGPSWatshes2019}.} as well as in military cyber-attacks affecting thousands of civilian ships~\cite{C4ADS2019}. Despite these limitations, there is active research on building secure and privacy-preserving localization solutions based on GPS by introducing additional protocol and security mechanisms, for instance, GPS-based active crowd localization based on digital signatures and bulletin boards applied for tracking lost items~\cite{Agadakos2017,Yucel2018}.

Mobile cellular network providers have been earlier proposed to act as oracles to submit positioning information to smart contracts that verify whether such positions are included into virtual borders referred to as \emph{geofences}~\cite{Victor2018}. Such geofences are represented by location encoding systems, for instance, Geohash and S2, that are hierarchical, i.e. they can model different cells at different resolution level. A geofence can be used by a local community to self-regulate its (i) decision-making territory and (ii) crypto-economic activity resulting from the incentivized participation in decision-making. The former determines the validation territory of witness presence claims. The latter determines the geographic areas in which transactions are permitted with collected tokens. For instance, Platin aspires to support such crypto-currencies for humanitarian aid use cases~\cite{Platin2019}. To control transaction costs for the execution of smart contracts, localization can be performed with different schemes: at regular time or distance intervals, on demand or upon violation of a citizen's presence in a geofence. Localization via mobile cellular networks can only though take place within the covered area of the mobile operator and global coverage requires special roaming service and collaboration between different mobile network operators. An alternative approach to overcome this limitation is to allow cellular towers of any mobile network to provide secure location services for the blockchain. Such an approach is earlier introduced. It involves cellular towers with a well defined location that issue location certificates and participate in mining location proofs. Trust is achieved using cryptographically signed IP packets~\cite{Dasu2018}.

An alternative infra\-stru\-ctu\-re to the proprietary and closed networks of mobile operators is the use of Low Power Wide Area Networks that allow access to an unlicensed radio spectrum~\cite{Raza2017}. LPWAN provide the following alternative trade-offs: long range, low power operation at the expense of low data rate and high latency. For instance, The Things Network~\cite{TheThingsNetwork2019} builds a global open LoRaWAN network of 7231 gateways in 137 cities run by local self-orga\-ni\-zed communities providing extensive coverage in urban environments. FOAM intends to use this decentralized open infrastructure for secure location verification enforced by smart contract safety deposits. Proof of location is performed within a \emph{zone} (community operator) defined by at least four \emph{zone authorities} (radio gateways) each managing a number of \emph{zone anchors} (radio beacons). A zone anchor is a device with a radio transmitter, a local clock and a public key. It is capable of engaging in a Byzantine fault-tolerant clock synchronization protocol~\cite{Malekpour2015}. Zone anchors perform triangulations and verify claims of presence via authentication certificates that are fraud proof. A zone authority is a node with an Internet connection that determines whether the zone anchors are in sync. 

All of the above solutions among others~\cite{Javali2016,Khan2014,Li2016} require additional special infrastructure. Mobile cellular networks and LPWAN may be unavailable or underperformimg in cases of natural disasters and unpredictable high-density mobility patters. In these scenarios, an alternative infra\-stru\-ctu\-re-inde\-pe\-ndent and decentralized approach is the use of peer-to-peer ad hoc (opportunistic) networks formed by self-o\-rga\-ni\-zed citizens' devices running decentralized secure protocols based on blockchain proof of stake consensus mechanisms~\cite{Amoretti2018}. Proofs of location are performed between \emph{witnesses} and a \emph{prover}, whose Bluetooth interactions verify the identities of the involved devices as well as whether the location claims of each device are reachable within the radio coverage supported by the communication technology of the devices. Spatio-temporal mobility patterns of users may influence the verification process and additional measures of verification may be required, for instance, analysis of betweeness in pseudonym correlation graphs~\cite{Zhu2013} or social tracking distance metrics~\cite{Yucel2018}. Periodically changing the device identifiers according to a Poisson distribution prevents the reveal of real identities by observing location proof records~\cite{Zhu2013}. 

\subsection{Situation awareness and proving witnessing}\label{subsec:witness-presence}

A few blockchain approaches combine network-based with social-based proof of location~\cite{Lyu2015,Platin2019,Wolberger2018}. For instance, on-chain location claims at Platin consist of a public key and a proof of correctness. In practice this is the output of one out several locally executed algorithms that validate location information based on the following three security pillars: (i) \emph{sensor fusion}, (ii) \emph{behavior over time} and (iii) \emph{peer-to-peer witnessing}. Sensor fusion relies on multiple sources of sensor data, i.e. GPS, wireless access points, cell tower and Bluetooth oracles, for validation of location claims. Behavior over time reasons about any behavioral anomaly that indicates spoofing. Data-driven verification can be localized to preserve privacy by design and prevent turning proofs of witness presence to surveillance actions that can actually undermine and manipulate democratic processes~\cite{Haggerty2010}. Peer-to-peer witnessing using ad hoc opportunistic networks can be used as an additional counter-measure to testify for attackers that may replay sensor fusion or report fake behavior over time. 

Proofs of witness presence verify the situation awareness required for a more informed collective decision-making. For instance, assume a crowd-sensing collective movement for a spatio-temporal safety assessment of bike riding in a city. Citizens rate the safety of different points of interests in the city based on which new data-driven policies can be designed to encourage the further safe use of bikes and the improvement of the infrastructure, i.e. new bike lanes. Making safety rating on the points of interest subject of proving witness presence can potentially improve the rating quality and as a result the effectiveness of a new designed policy. Beyond citizens proving their location, proving bike riding experience, on spot or elsewhere, indicates a situation awareness with an added value and a higher potential for a more effective policy. Verification can be performed on-chain or off-chain using witnesses, sensor fusion, i.e. analysis of GPS/accelerometer data, or even oracles, i.e. a bike sharing operator. 

Other means to verify witness presence include the following: Contextual QR codes~\cite{Rouillard2008}, challenge questions, puzzles and CAPTCHA-like tests~\cite{SIKORKA2019}, whose solutions require information mined at the point of interests. In addition, collaborative social challenges~\cite{Beckers2016,Aladawy2018} between citizens are means to introduce social proofs based on social psychology as well as community trust for protection against social engineering attacks~\cite{Schaab2017}. Moreover, communities can also institutionalize their own digital witnesses based on privacy-preserving forensic techniques introduced in the context of blockchain~\cite{Ugwu2018,Nieto2018}.

\subsection{Blockchain and consensus network}\label{subsec:blockchain}

Figure~\ref{fig:blockchain} illustrates the blockchain-based Internet of Things architecture with which witness presence claims are verified. The architecture is a layered one, starting from the physical public space where localization in points of interest is performed by wireless beacons using solutions such as the ones reviewed in Table~\ref{table:proof-of-location}. Proofs of location can be augmented with one or more layers of social proofs using methods outlined in Section~\ref{subsec:witness-presence}. Full nodes with computational power and an Internet connection participate in the consensus network to further verify and cross-check the adherence to protocol rules across the local nodes at each point of interest. Verified witness presence claims are written to the blockchain. They are a result of location proofs, social proofs and protocol adherence proofs performed over the layered architecture. 

\begin{figure}
\centering	
\includegraphics[width=0.9\columnwidth]{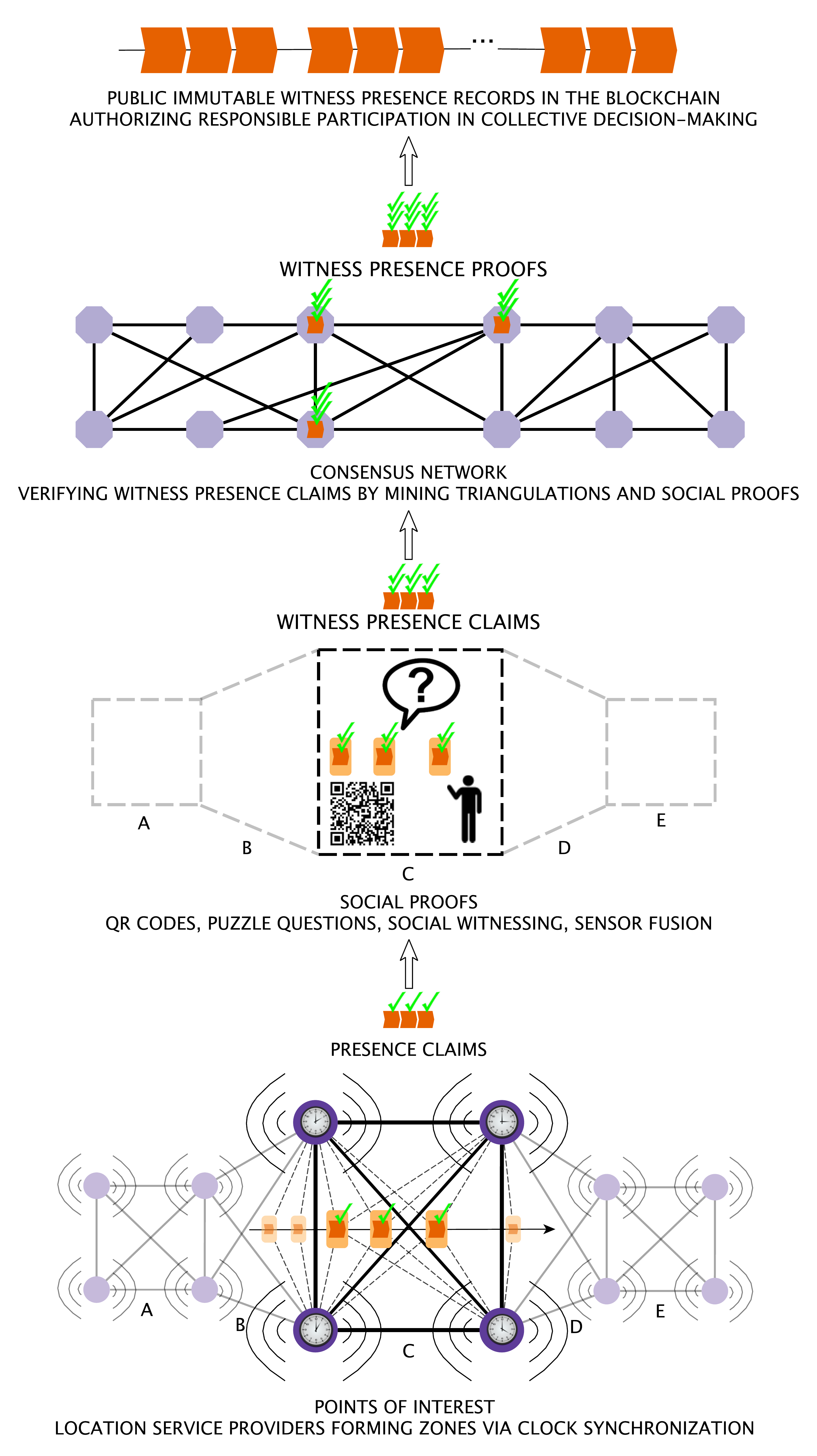}
\caption{A blockchain-based Internet of Thi\-ngs architecture for proving witness presence. Points of interest in an urban physical space can be determined by the transmission coverage zone of wireless beacons that act as secure location service providers using triangulation and Byzantine fault-tolerant clock synchronization~\cite{Malekpour2015}. Presence claims can be further supported by social proofs on spot that verify the situation awareness of citizens in collective decision-making. Witness presence claims are further verified in a blockchain consensus network that consists of (full) nodes with Internet connection. They verify whether the rules for location and social proofs are fulfilled. Location accuracy can be traced and checks for fraud can be performed, for instance, comparing location claims from different adjacent points of interest to verify whether clocks are actually in sync. Verified witness presence claims are finally written to the blockchain based on which a more responsible participation in collective decision-making can be authorized.}\label{fig:blockchain}
\end{figure}

The properties of blockchain consensus for proving witness presence are outlined as follows: (i) \emph{Validator set}: The validators of presence claims depend on the adopted approach from Table~\ref{table:proof-of-location}. For instance, approaches such as LPWAN and Peer-to-peer (P2P) Ad Hoc networks that rely on distributed networks of wireless beacons determine their validator set based on their physical distance. Communication constrained by physics result allows the validators in close physical proximity to verify location claims around a point of interest~\cite{Jiang2020}. This set of validators can be further expanded with nodes for proof of stake. Such nodes hold a public key and stake a deposit token to validate social proofs. (ii) \emph{Validator weight}: The number of staked tokens can be used as a weight. However, other (reputation) criteria related to the level of participation and democracy could be engaged~\cite{Weatherford1992,Levi2009,Etter2018}: to what extent a geographic region decides public matters via witness presence, the level of legitimacy of witness presence in a region, and other. (iii) \emph{Validator criteria}: Proof of work solves a cryptographic puzzle that verifies the validity of a block (its \texttt{nonce} number) when its hash value is lower than a difficulty threshold: \texttt{sha(nonce)<difficulty}. In contrast, proof of witness presence requires matching the signature to the validator set, meeting the minimum stake requirement and having no slashing conditions, e.g. Byzantine fault-tolerant clock synchronization is successfully performed for proving pre\-se\-nce claims~\cite{Malekpour2015}. Verification rules for robust spatio-temporal data can be further engaged here~\cite{Nasrulin2018,Bornholdt2019}. (iv) \emph{Validator verifiability}: For presence claims, signed receipts of all clock synchronization messages received and synced to the chain are required. The limits of transmission coverage restrict the receipt of such messages from validators within the proximity of a point of interest~\cite{Jiang2020}. Social proofs require keeping the chain synced to verify that other validators have staked and belong to the validator set. 

In terms of the crypto-economic incentive model, a utility token~\cite{Ballandies2018} can be used to reward (i) citizens and communities for introducing localization infrastructure for location proofs, (ii) the establishment of social proofs in points of interests for proving social claims, or (iii) the use computational resources for validation of the witness presence claims in the consensus network. The rewards include minted new tokens and transaction fees according to the protocol rules enforced by the network itself that punishes adverse behavior. In all these cases, permissionless participation requires staking that is the commit of a deposit token value, while faults resulting in violations of the protocol rules (slashing conditions) result in penalties. These are usually magnitudes higher than the anticipated short term rewards. Therefore, the \emph{entry cost}, \emph{existence cost} and \emph{exit penalty} can make proofs of witness presence resistant to Sybil attacks~\cite{Otte2017,Nasrulin2018,Bornholdt2019}. Note that citizens who make witness presence claims require to pay a fee to witness presence service providers of the local community\footnote{These are the nodes performing the localization and the social proofs. Therefore, no service fee needs to be payed to a central authority.} in the same utility token, another token or fiat money. These fees reward the further development and maintenance of the infrastructure, i.e. supporting witness presence in new points of interest, improving the localization accuracy, increasing the bandwidth allocation, augmenting further the points of interest with social proofs, etc. Citizens may have a self-interest to reward such participatory processes directly from their own funds as the means to improve direct democracy and give themselves a stronger voice on public matters. Such funds may also originate by state authorities incentivized to improve the legitimacy of collective decision-making in the same way that such funds are reserved for conducting elections, e.g. running voting centers. In other words, witness presence turns points of interests into are a new type of digital voting centers for augmented decision-making available at any time and location. 

The transaction costs of proving witness presence claims are dependent on mobility patterns and the density of the witness presence claims made by citizens at each point of interest. They also depend on the available radio beacons covering a point of interest as such devices have physical constraints on the rate of messages they can process. The feasibility of permissionless Byzantine consensus protocols to operate in real-time over wireless networks is recently demonstrated~\cite{Jiang2020}. Benchmark measurements of transaction latency are available in earlier work based on which the choice of inter-block time, the number of confirmation blocks and process-level changes can be tuned~\cite{Yasaweerasinghelage2017}. Smart contracts can be designed to load-balance transaction costs between location proofs and social proofs: within a large crowd concentrated on a point of interest, social proofs may prove to be more reliable that location proofs made by overloaded radio beacons. Moreover, further performance improvements can be achieved via a hierarchical Plasma design that splits the blockchain into parent-child chains~\cite{Poon2017,Ziegler2019,FOAMResearch2019}. A child chain is constructed for each point of interest running synchronous consensus for clock synchronization. In contrast, a parent chain holds the staked tokens and the smart contracts that represent the different child chains. The parent chain may rely on an asynchronous consensus network in  Ethereum such as Nakamoto in the case of proof of work or Casper in case of proof of stake~\cite{Lao2020,FOAMResearch2019,Ballandies2018}. 

A self-sovereign identity management system~\cite{Norta2019} can be used to authenticate citizens' actions in the proposed permissionless distributed ledger, i.e. verifying the actual citizen who issued a witness presence claim to prevent double participation that can influence the result of collective decisions~\cite{Kuperberg2019}. Moreover, the information provided to the smart contracts for social proofs can be further used for multi-factor authentication~\cite{Csahan2019,Abayomi2019,Taher2019}. Identity management services do not need to rely on third parties and several such services are earlier proposed and reviewed~\cite{Lim2018}. In particular,  UniquID~\cite{UniquID2020} is an identity and access management service for the Internet of Things that is open-source, permissionless and relies on Ethereum~\cite{Lim2018}. LifeID is another self-sovereign digital identity platform with which citizens control all transactions that require authentication of their identity without the need for third-party corporations or government agencies. Zero-knowledge proofs are applied and the minimum data required for verification are shared~\cite{Bokkem2019}.

%\cite{Sullivan2017} e-identity in estonia

%\cite{Kuperberg2019}
%
%The contribution of this paper is a thorough analysis of the opportunities and state-of-the-art at the intersection of blockchain technology and government-issued electronical identity documents (eIDs), including existing implementations and pilots.

%\cite{Falcone2019}
%Manufacturing

\section{Real-time Collective Measurements}\label{sec:collective-measurements}

Real-time collective measurements are the aggregation of citizens' crowd-sensing data, e.g. decisions, made as a result of witness presence. The computation of aggregation functions, e.g. summation, mean, max, min, standard deviation, are some examples of such collective measurements. They can be used as follows: Citizens receive real-time crowd-sensing information. A collective awareness is built that is used as live feedback for future crowd-sensing decisions, i.e. the feedback loop in Figure~\ref{fig:architecture}. Collective measurements may encourage or discourage witness presence, for instance, a warning system that guides authorities to mitigate a physical disaster in certain points of interest, while citizens are instructed to avoid dangerous ones. 

A transparent and reliable system for collective measurements is paramount for building collective awareness and trust among citizens, both required for a viable augmented democracy paradigm. Existing centralized polls and social media often fail to provide reliable and trustworthy information and are often subject of citizens' profiling over collected personal data, nudging and political manipulation~\cite{Fisher2018,Vosoughi2018,Lazer2018}. Instead, the computations required for aggregation can be crowd-sourced to citizens using their personal devices or computational resources of communities in a similar fashion as the diaspora* social network~\cite{Bielenberg2012} or Scuttlebutt~\cite{Scuttlebutt2019,Tarr2019}. Although decentralized computations for aggregation are more privacy-preserving by design using differential privacy and homomorphic encryption techniques, their accuracy requires significant self-adaptations to cope with the following: (i) continuous data streams as a result of changes in decision-making, (ii) a varying spatio-temporal participation level as well as (iii) (Byzantine) failures. 

The relevance of these challenges in the augmented democracy paradigm is the following: Citizens revisiting a point of interest in the future may reevaluate an urban quality triggering recomputations of the aggregation functions to reflect changes on the input crowd-sensing data. The decision of a citizen updates the aggregation functions as long as witness presence is proved. If witness presence cannot be verified anymore, corrective rollback operations on the aggregation functions are performed to reflect the latest status of participation. Similarly, any failure that cannot guarantee a correct execution of the aggregation protocol shall be treated as a failure to verify witness presence and therefore, corrective operations with rollback operations are performed in this case as well. In summary: \emph{collective measurements provide a live pulse of a crowd, whose localization at points of interest is verified for witness presence}. 

A possible feasible decentralized approach to realize this ambitious concept is the use of DIAS, the \emph{Dynamic Intelligent Aggregation Service}~\cite{DIAS2019,Pournaras2017c}. DIAS is a network of interconnected agents deployed in citizens' personal devices or in computational resources of regional communities around points of interest. Agents perform a gossip-based communication to disseminate crowd-sensing data used as input in aggregation functions computed locally by each agent. The agents of DIAS are self-adaptive and can update the aggregates in an automated way when input data change as well as when agents join, leave or fail~\cite{Pournaras2017e,Pournaras2017f}. They have this capability by reasoning based on historic data in a privacy-preserving way. Reasoning relies on a distributed memory system that consists of probabilistic data structures, the Bloom filters~\cite{Pournaras2017c}. In simple words and practical terms, the memory system can reason whether the choice a citizen has changed at a point of interest. It can also reason on whether a citizen visits again or leaves a point of interest. Further technical information about DIAS is out of the scope of this paper and readers are referred to earlier work~\cite{Pournaras2017c,Pournaras2017e,Pournaras2017f}. 

Collective measurements can be made conditional to different witness presence scenarios that are referred to as \emph{collective measurements maps}. Two types of such measurements maps are introduced as an illustrative example:  (i) \emph{distributed} and (ii) \emph{localized}.

In the distributed measurements maps, aggregation functions receive the input data of citizens, who prove witness presence in one out of several possible points of interest. In other words, a logical disjunction (OR) determines the proof of witness presence at one possible point of interest as the required condition to participate in the collective measurements. This measurements map is relevant for federated democratic processes of regional communities, for instance collective decision-making in the spatial context of multiple university campuses, i.e. an `eduroam' version of augmented democracy. Figure~\ref{fig:users-poi}-\ref{fig:aggregation} illustrate the augmented democracy paradigm with a distributed measurements map.

\begin{figure*}
\centering	
\includegraphics[width=1.0\textwidth]{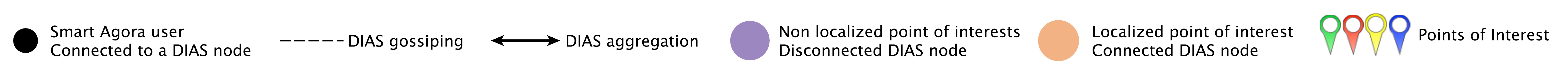}\\
\subfloat[A snapshot of citizens moving around with their smart phones to visit augmented points of interest.]{\includegraphics[width=1.0\columnwidth]{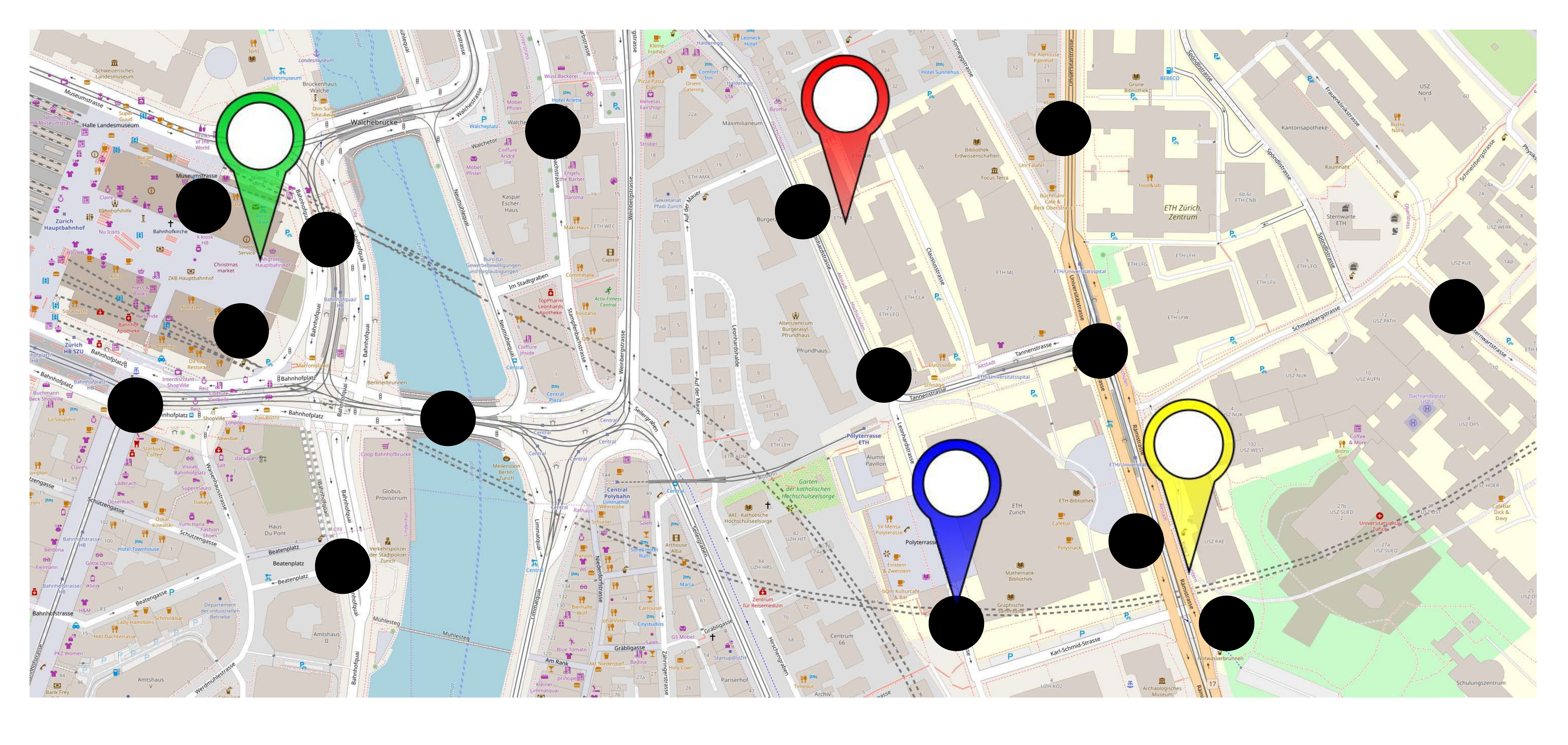}\label{fig:users-poi}}\hfill
\subfloat[Each point of interest has a verified number of citizens proving their witness presence.]{\includegraphics[width=1.0\columnwidth]{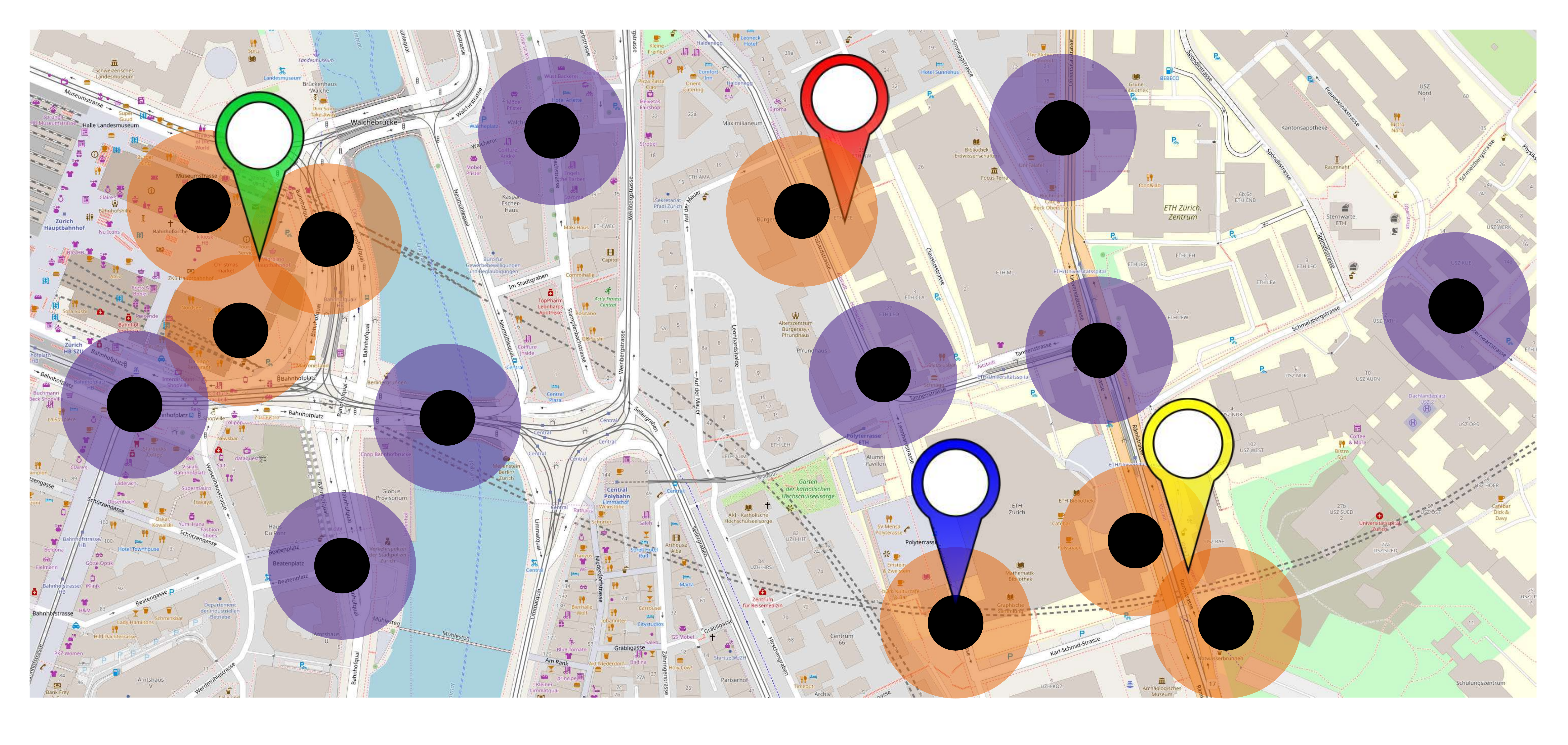}\label{fig:localization}}\\
\subfloat[Citizens are interconnected in a decentralized network of gossip-based communication over which collective measurements, i.e. data aggregation, can be performed.]{\includegraphics[width=1.0\columnwidth]{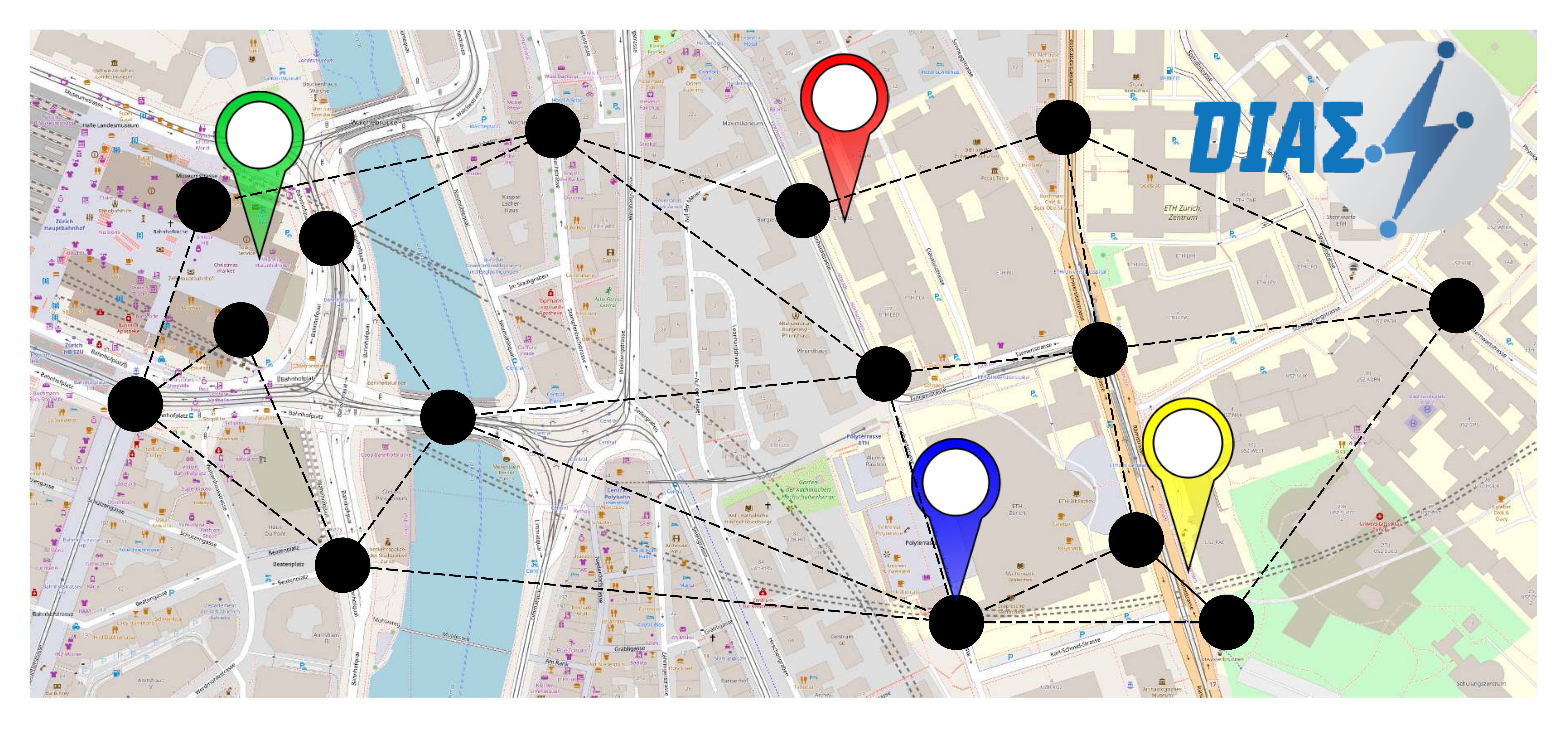}\label{fig:DIAS}}\hfill
\subfloat[Collective measurements are exclusively performed between the citizens with a proof of witness presence.]{\includegraphics[width=1.0\columnwidth]{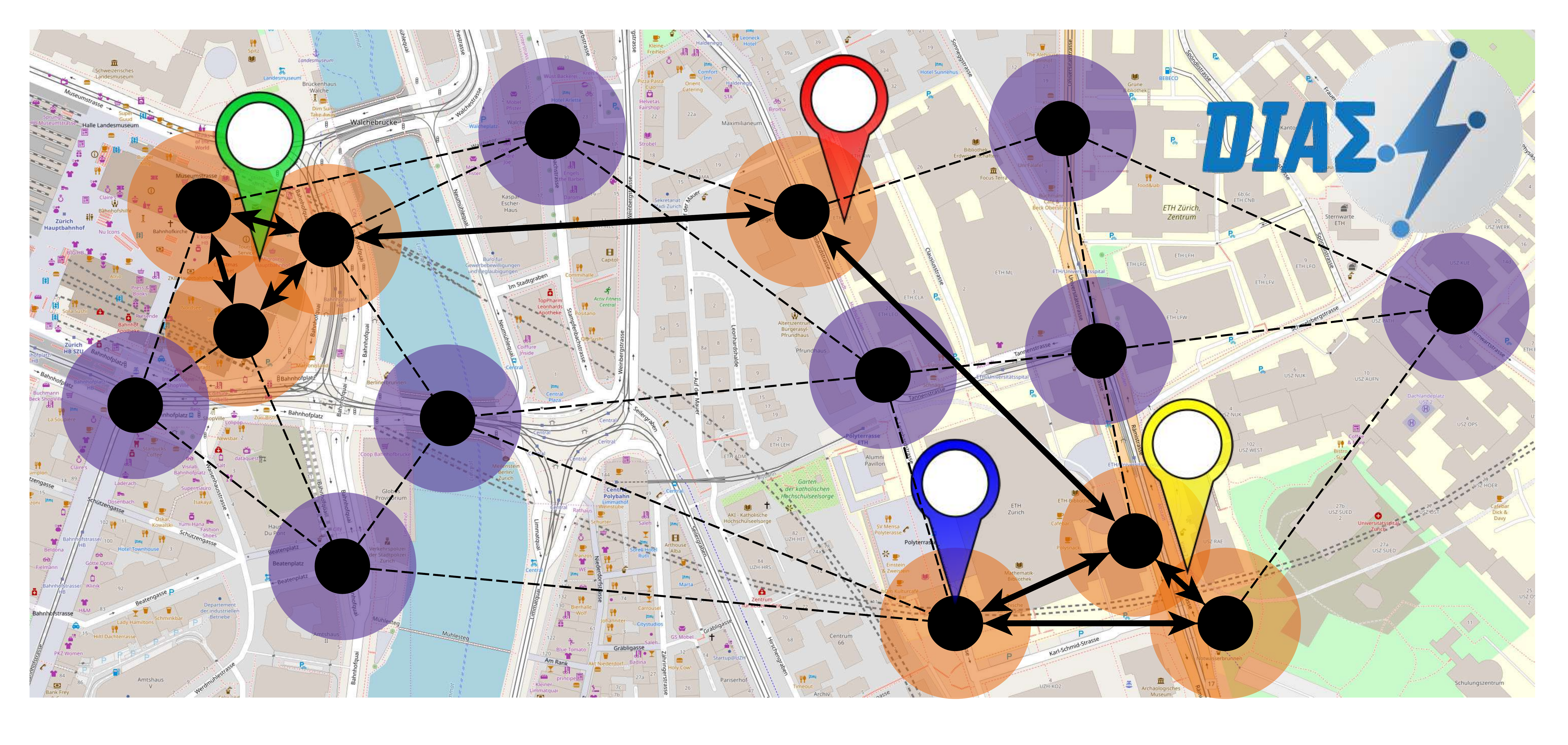}\label{fig:aggregation}}\\
\subfloat[Regional community A]{\includegraphics[width=0.49\columnwidth]{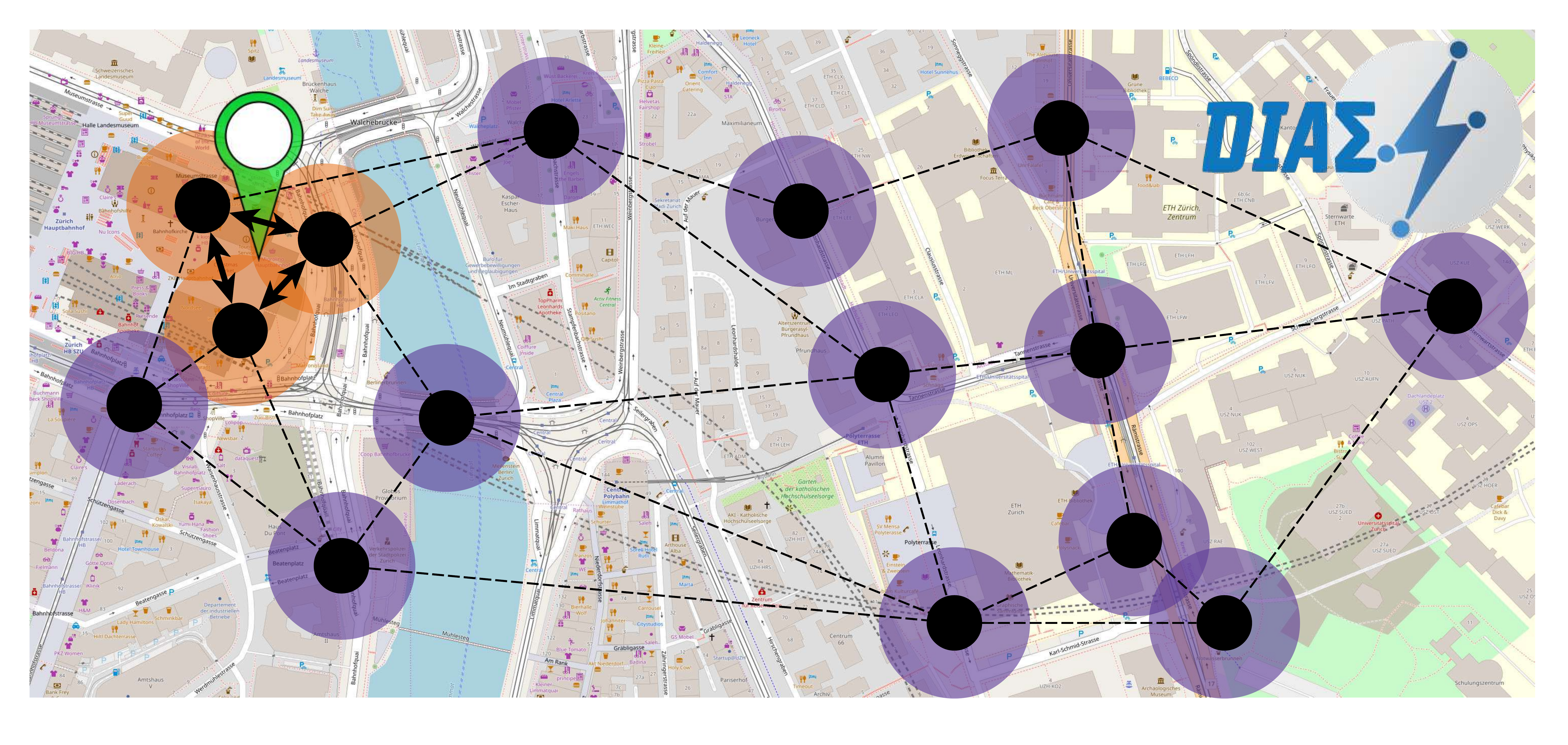}\label{fig:network-a}}\hfill
\subfloat[Regional community B]{\includegraphics[width=0.49\columnwidth]{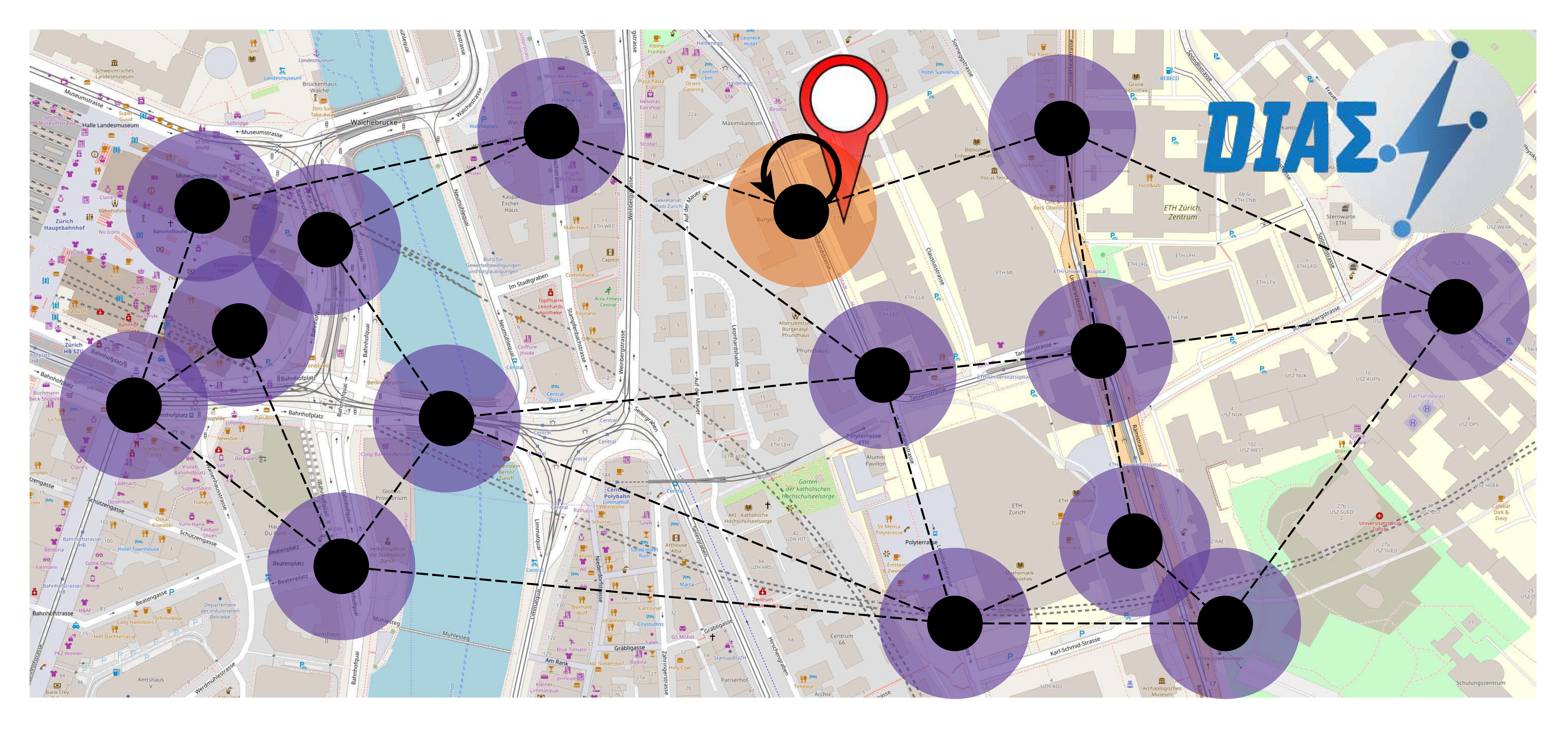}\label{fig:network-b}}\hfill
\subfloat[Regional community C]{\includegraphics[width=0.49\columnwidth]{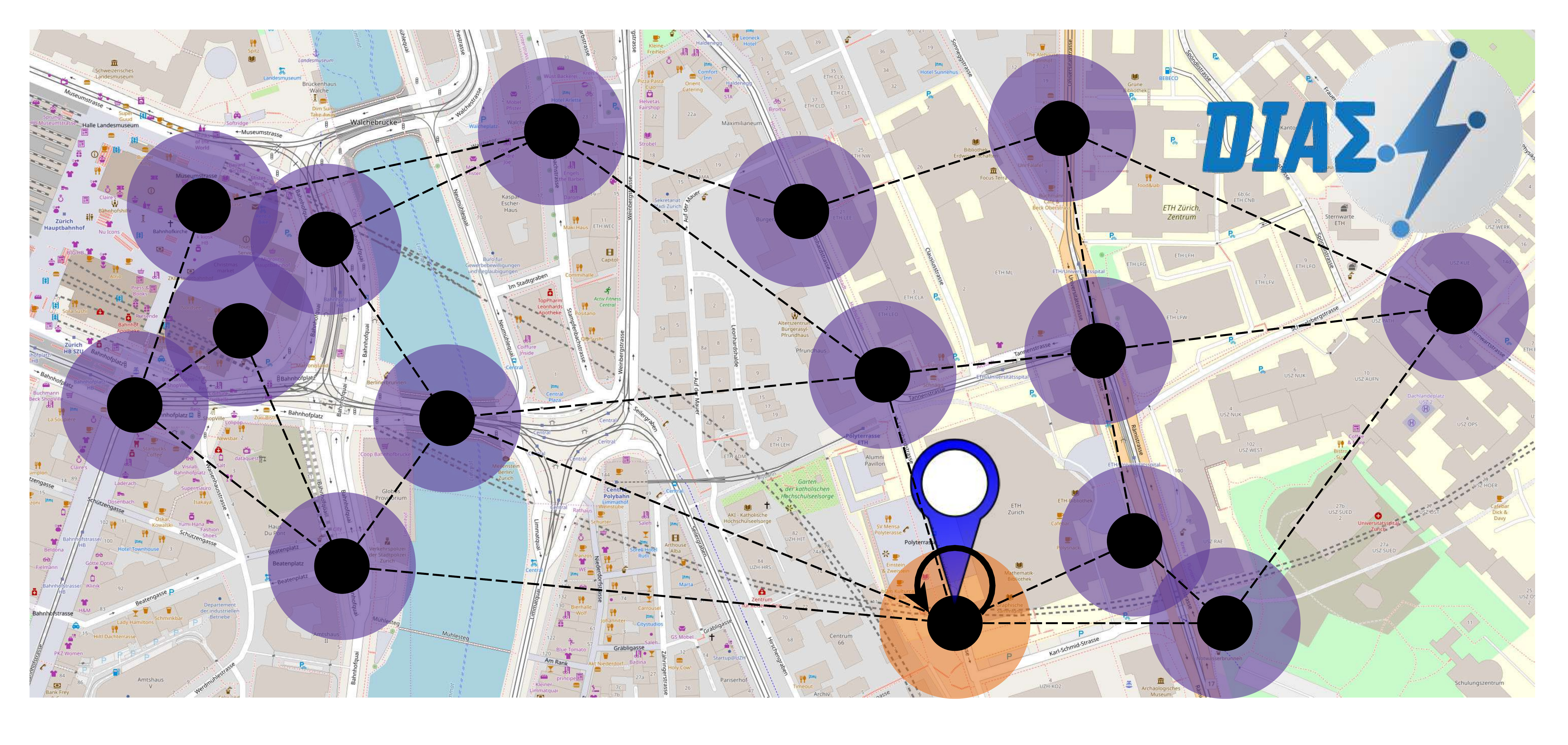}\label{fig:network-c}}\hfill
\subfloat[Regional community D]{\includegraphics[width=0.49\columnwidth]{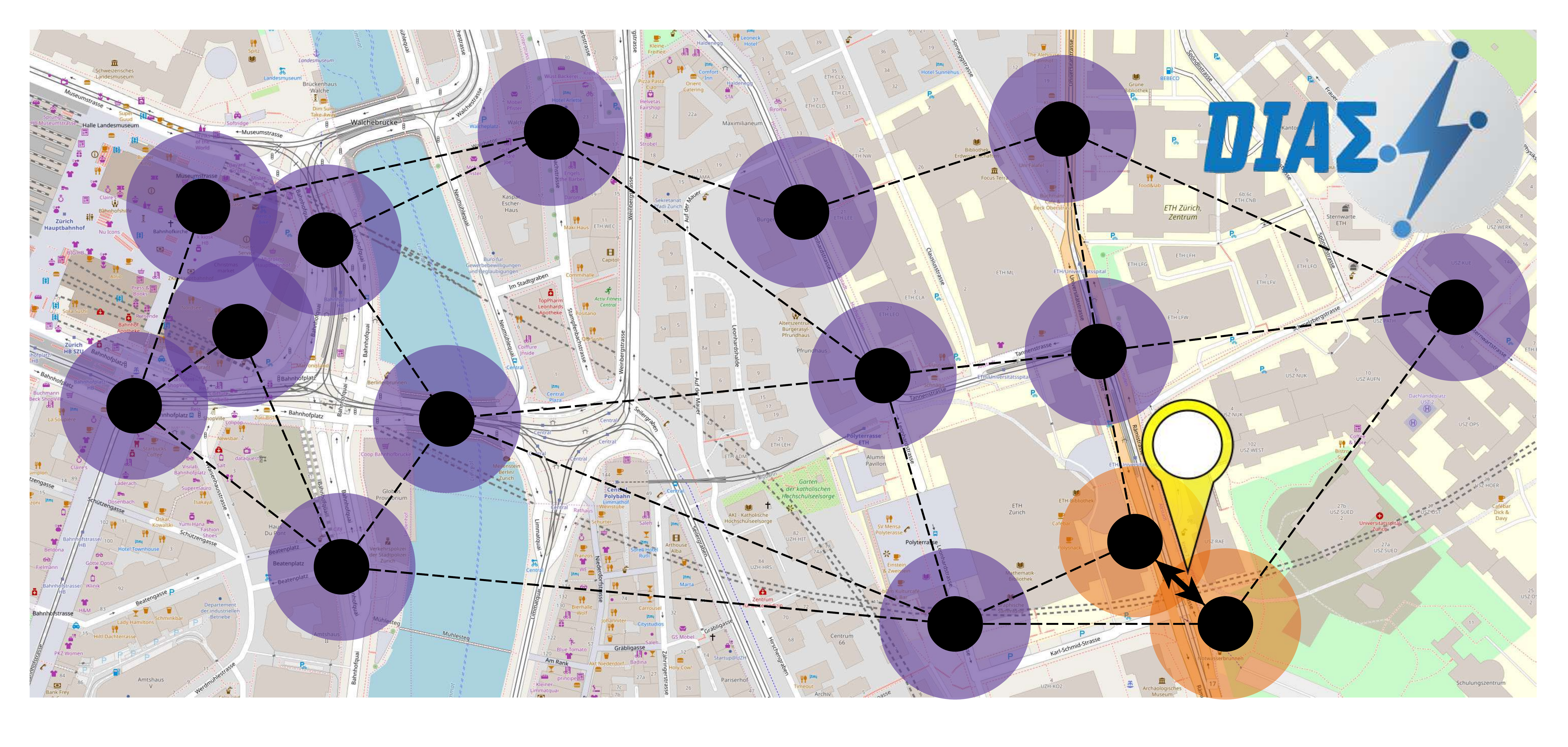}\label{fig:network-d}}\hfill
\caption{An illustration of the augmented democracy paradigm. \textbf{Distributed measurements map in Figure~\ref{fig:users-poi}-\ref{fig:aggregation}}: Collective measurements are performed by proving witness presence at one out of several possible points of interest. \textbf{Localized measurements map in Figure~\ref{fig:network-a}-\ref{fig:network-d}}: multiple localized collective measurements are performed by proving witness presence at a certain point of interest.}\label{fig:encapsulations}
\end{figure*}

In localized measurements map, aggregation functions receive citizens' input data by proving witness presence at a certain point of interest. This measurements map is relevant for local regional communities that use their own computational resources to run their own collective measurements and make them available to their local citizens. Figure~\ref{fig:network-a}-\ref{fig:network-d} show an example. For each point of interest, aggregation is restricted between the localized citizens proving witness presence. 

The two proposed collective measurements maps are not the only options and more complex witness presence logic can be designed. For instance, semantic collective measurements can run by two DIAS networks aggregating crowd-sensing data at points of interest corresponding to (i) tram stations and (ii) bus stations respectively. 

The communication complexity of such real-time collective measurements exclusively depends on the updates of the input data in the aggregation functions. Such updates are triggered by (i) changes of the input data and (ii) join and leaves of nodes in the network that result in new input data or data removals. The influence of such updates in the aggregation accuracy is studied in earlier work~\cite{Pournaras2017c,Pournaras2017e,Pournaras2017f}. In the augmented democracy paradigm, the following factors influence the trigger of such updates: (i) A higher number and density of the points of interest in which witness presence can be verified (joins/leaves) is likely to cause a higher number of input data updates and as a result higher communication cost. This is especially the case for the distributed measurements maps. (ii) The citizens' mobility patterns over the points of interests. More frequent witness presence claims in the different points of interest result in higher communication cost.

\section{Evaluation Methodology and Results}\label{sec:evaluation}

Evaluating the end-to-end integrated functionality of the whole augmented democracy paradigm illustrated in Section~\ref{sec:paradigm} is a challenging endeavor. This requires a rigorous extensive evaluation of each proposed pillar that is subject of active ongoing work~\cite{Pournaras2019e}. Such detailed evaluation does not fall within the scope and objectives of this paper. To overcome the aforementioned challenge and come with a very first proof of concept, a simple yet fully-fleshed experimental testnet scenario is designed with the following requirements: (i) A realistic Smart City use case for participatory crowd-sensing. (ii) Proof of witness presence in two points of interest based on GPS. (iii) Real-time collective measurements in distributed measurements maps over a small crowd of test users with different realistic mobility patterns. 

Moreover, the quality of information collected based on citizens' witness presence is validated using empirical official data from public authorities. More specifically, an application scenario of cycling safety in Zurich is studied, in which the perception of bike riders about the cycling safety in different urban spots is compared to an empirical safety model built using official data of the Federal Roads Office collected from Swiss GeoAdmin~\cite{SwissGeoAdmin2019,Castells2019}. If the two safety estimations match, then this is indication that witness presence in participatory crowd-sensing can indeed provide information quality comparable to the official but costly data collection methods.

\subsection{Experimental testnet scenario}\label{subsec:testnet}

A testnet scenario on sustainable transport usage is introduced to address the first requirement for a proof of concept. The testnet scenario ran for about one hour on 3.6.2019 between 13:00-14:00 in Zurich. The goal of the testnet scenario is to assess the preferred transport mean with which citizens visit a place they witness. Such a use case is relevant to transport engineers, who work with travel diaries. While travel diaries are modeled based on traditional, costly and infrequent survey questions, the pervasiveness of the Internet of Things promises new opportunities for more realistic and real-time data collection based on which future traffic flow models can rely on~\cite{Danaf2019,Prelipcean2018}. Similarly, city councils can establish new policies and incentives for citizens to make use of more sustainable transport means. 

This use case assumes a linear model of sustainability over six transport means: 0. \emph{Car}, 1. \emph{Bus}, 2. \emph{Train}, 3. \emph{Tram}, 4. \emph{Bike}, 5. \emph{Walking}. These transport means are common in Zurich and usually a destination can be reached fast with several different transport means. Car comes with the minimum sustainability value of zero, while walking comes with the maximum sustainability of 5. Although this linear model is an oversimplification over several involved sustainability aspects such as environment, health, safety, social and other, it is intuitive and straightforward to engage test users as well as interpretable. Therefore, the purpose of the use case is to serve the realism of the testnet scenario rather than collecting use case data for a rigorous analysis. 

The second requirement is met by designing a decision-making process in Smart Agora for the testnet scenario. The test users make a choice via a likert scale question that pops up in the Smart Agora app when they are localized at a point of interest as shown in Figure~\ref{fig:app-localization}. Such a question is part of six crowd-sensing Smart Agora assets created for six test users, who are equally split into two groups. 

%NEW TEST
\begin{figure}
\centering	
\subfloat[Localization.]{\includegraphics[width=0.49\columnwidth]{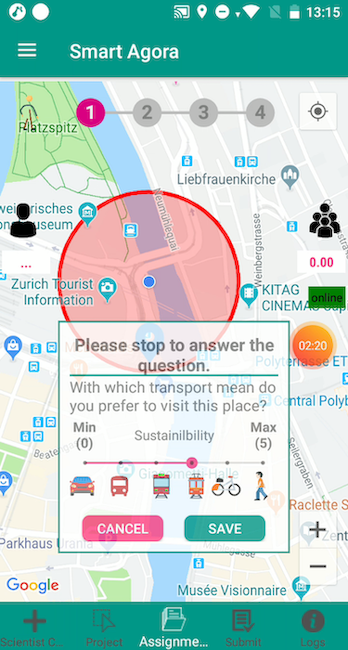}\label{fig:app-localization}}\hfill
\subfloat[Aggregation.]{\includegraphics[width=0.49\columnwidth]{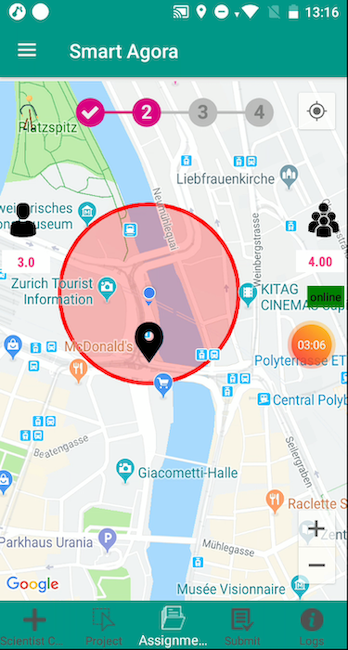}\label{fig:app-aggregation}}\hfill
\caption{Assessing the preferred transport mean to reach a witnessed point of interest in terms of sustainability. Localization triggers a question followed by live collective measuments received from other test users localized to other points of interest.}\label{fig:app-testnet}
\end{figure}

To meet the third requirement, each crowd-sensing asset is designed in the sequential navigational modality with two points of interest visited in reversed order among the two groups to assess the distributed measurements maps of DIAS, i.e. choices of test users are aggregated in real-time from two different remote points of interest. Figure~\ref{fig:scenario} illustrates the designed experimental scenario. Note that the depicted walking path is the calculated Google Maps path rather than the one that test users followed\footnote{Group 1 has followed a shortcut on the way to ETH Zurich Hauptgeba{\"u}de by using the Polybahn~\cite{Polybahn2019}.}. The actual traces collected with Smart Agora within the localization circles are shown in Figure~\ref{fig:GPS-traces} of Appendix~\ref{sec:traces}. 

%OLD TEST
%Orange, Group 1: 1 (50m), 2 (150m), 3 (100m)
%Purple, Group 2: 4 (100m), 5 (100m), 6 (150m)
\begin{figure}
\centering	
\includegraphics[width=1.0\columnwidth]{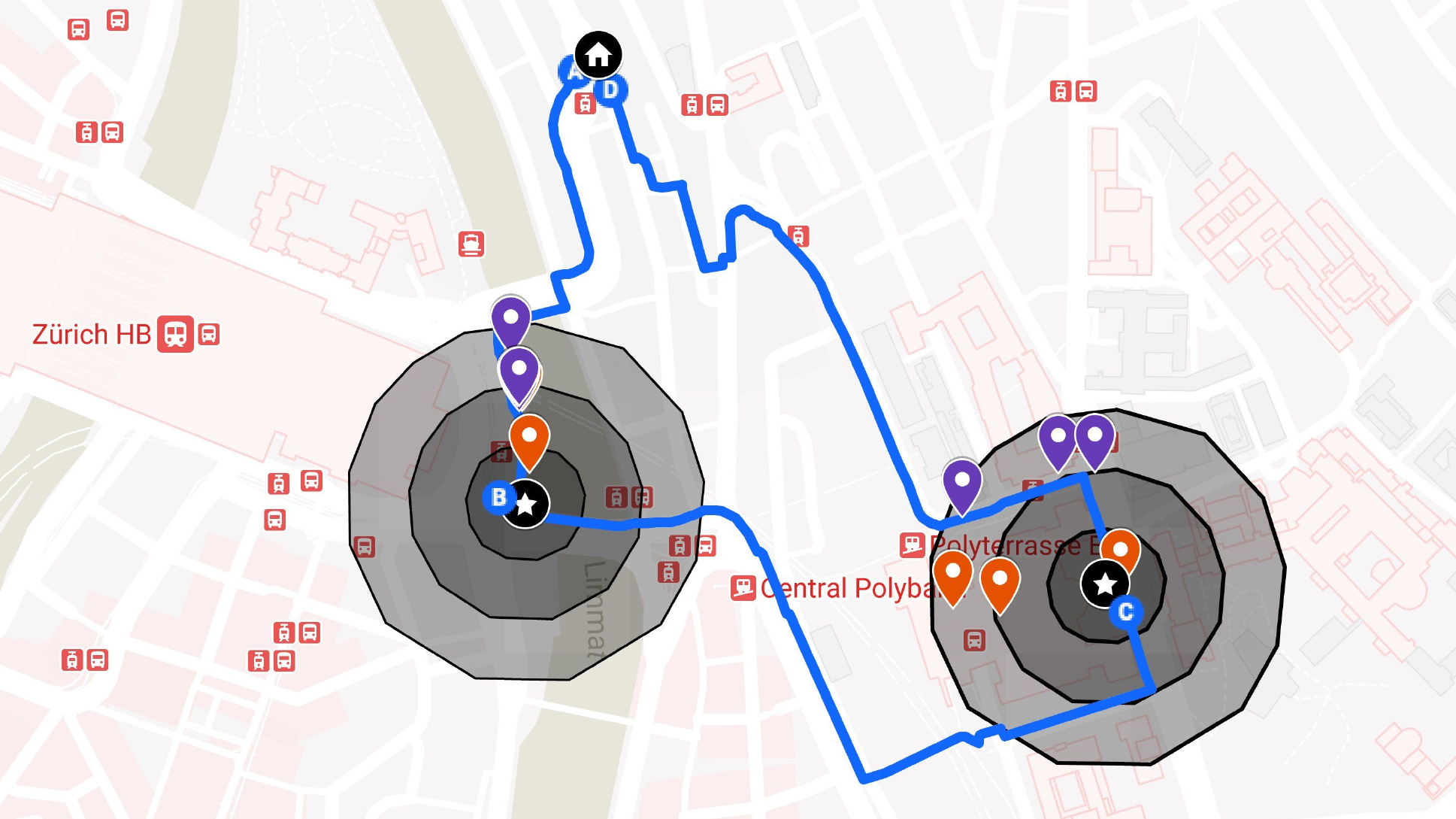}\label{fig:testnet}
\caption{An overview of the testnet scenario: Two groups each with three test users visit in reversed order the two point of interests of (i) Zurich Hauptbahnhof and (ii) ETH Zurich Hauptgeba{\"u}de starting from Stampfenbachstrasse 48, 8092, Zurich, where the Chair of Computational Social Science of ETH Zurich is situated. Group 1 (Orange) visits first Zurich Hauptbahnhof and Group 2 (purple) visits first ETH Zurich Hauptgeba{\"u}de. Each unique localization to one of the points of interest triggers for a test user a question for assessing sustainable transport usage. While a test user remains localized, live collective measurements among all other localized test users are received. The three nested circles around each point of interest visualize the three different ranges of localization that each group member has: 50, 100 and 150 meters.}\label{fig:scenario}
\end{figure}

To make sure that multiple test users are localized simultaneously in different points of interest, a requirement to evaluate the distributed measurements maps, a common starting point is chosen, the building of the Chair of Computational Social Science at ETH Zurich, which falls in close proximity between the two points of interest: (i) \emph{Zurich Hauptbahnhof} that is the main station of the Zurich city center and (ii) \emph{ETH Zurich Hauptgeba{\"u}de} that is the main building of ETH Zurich. Both groups start their navigation at the same time, i.e. mimicking two swarms. This makes the participation of the test users in the experimental process simpler. However, this localization synchronicity is an undesirable experimental artifact as in reality mobility patterns differ among citizens. To limit the synchronicity effect, each user has a localization circle with different radius value: 50, 100 or 150 meters. The circle, instead of an ellipse, is used here for simplifying the analysis and interpretability of the localization traces. 

Figure~\ref{fig:testnet-aggregation} illustrates the accuracy of the collective measurements for each group and test user. The estimates of the average transport sustainability that each test user receives approximate well the actual values. Note that users with higher localization radius receive aggregate estimates earlier and they have a larger\footnote{Localization circles with lower size in which test users do not remain for enough time may result in missing the receipt of collective measurements as observed in the second group at the Zurich Hauptbahnhof point of interest.} time span during which the receive collective measurements. 

%OLD TEST BUT NORMALIZED VALUES TO THE NEW TEST
\begin{figure}
\centering	
\subfloat[Group 1.]{\includegraphics[width=1.0\columnwidth]{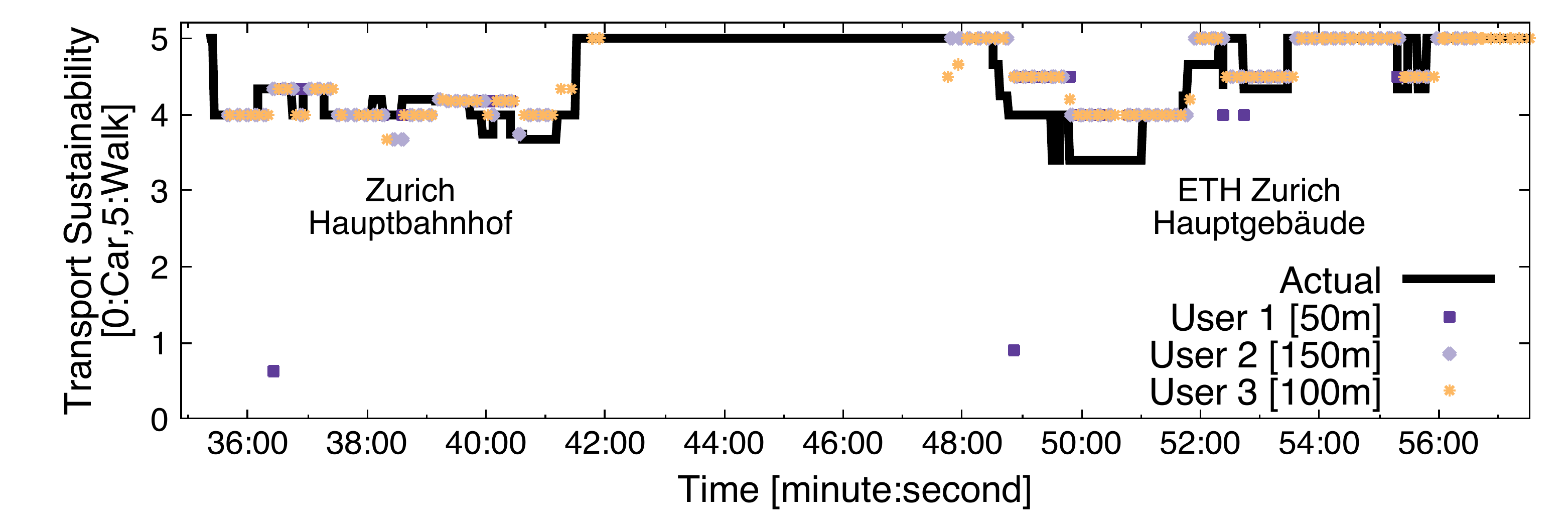}\label{fig:aggregation-group-01}}\hfill
\subfloat[Group 2.]{\includegraphics[width=1.0\columnwidth]{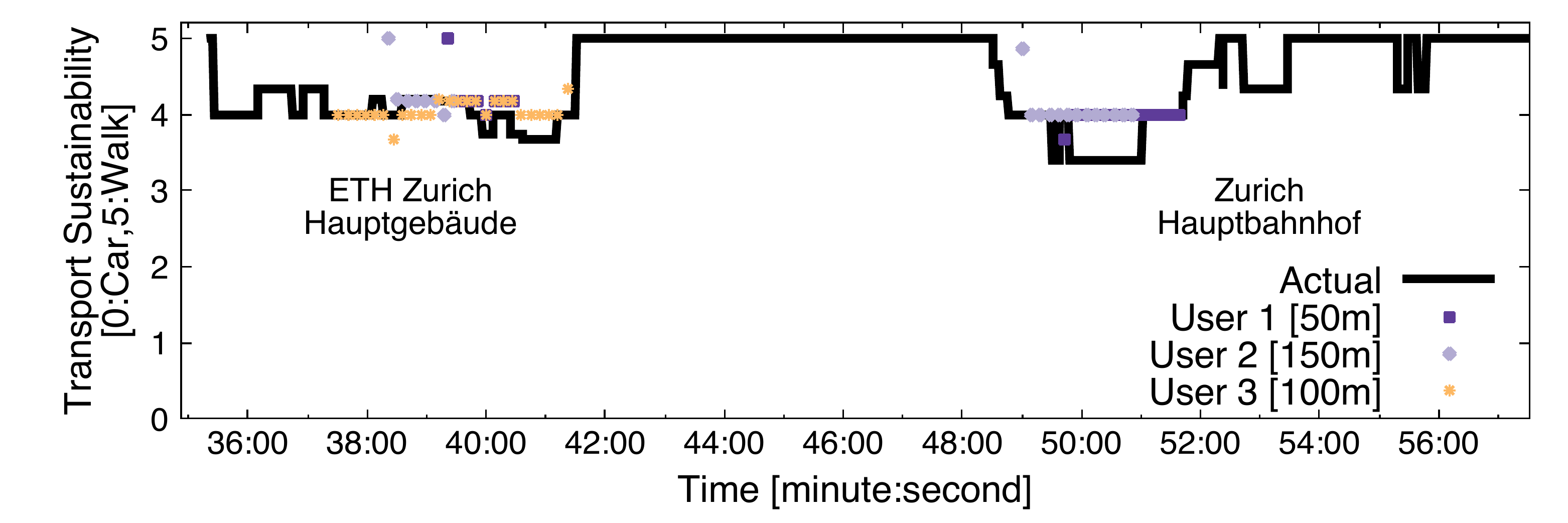}\label{fig:aggregation-group-02}}\hfill
\caption{Accuracy of real-time collective measurements during the testnet scenario on 3.6.2019 between 13:00 and 14:00 for 6 users split in 2 groups. The aggregation function calculated is the average transport sustainability among all test users localized in one of the two point of interests of Zurich Hauptbahnhof and ETH Zurich Hauptgeba{\"u}de.}\label{fig:testnet-aggregation}
\end{figure}	

Table~\ref{table:sustainability} shows the choices of transport means made by each test user at each point of interest. Overall, none of the more unsustainable transport means, i.e. car, bus and train, are chosen by test users to visit the points of interest. Walking and tram are the most popular means given that ETH Zurich and the main train station are very well connected with tram and are in close proximity with each other. The mean sustainability of 4.17 for ETH Zurich Hauptgeba{\"u}de is slightly higher than the one of 3.8 at Zurich Hauptbahnhof. 

\begin{table}
\caption{Transport sustainability responses for the two points of interest.}\label{table:sustainability}
\centering
\resizebox{\columnwidth}{!}{%
\begin{tabular}{l l l l}
\toprule
Group &  Test User  & Zurich Hauptbahnhof & ETH Zurich Hauptgeba{\"u}de \\\addlinespace\toprule
1 & 1 & 5. Walking & 3. Tram \\\midrule
1 & 2 & 3. Tram & 5. Walking \\\midrule
1 & 3 & 5. Walking & 5. Walking \\\addlinespace\toprule
2 & 1 & 3. Tram & 4. Bike \\\midrule
2 & 2 & 3. Tram & 5. Walking \\\midrule
2 & 3 & 4. Bike & 3. Tram \\\bottomrule\addlinespace
& Mean: & 3.8 & 4.17 \\\bottomrule
\end{tabular}
}
\end{table}

\subsection{Witness presence for cycling safety}\label{subsec:cycling-safety}

The cycling accident risk of the route in Figure~\ref{fig:cycling-path}b is studied that consist of four urban spots in Zurich. The risk estimation of this route is derived by a continuous spatial risk estimation model of the Zurich area that uses kernel density estimation with input the road network, geolocated accidents, their severity, and insurance compensation information~\cite{Castells2019}. The exact design of the model is out of the scope of this paper and the estimated risk values are used here as a baseline for comparison. In particular, this route is chosen for its extreme risk gradient observed around its circumference, with high risk at the top of the route and relative low/medium risk elsewhere as shown in Figure~\ref{fig:cycling-path}a. The actual risk values of the four urban spots are depicted in Figure~\ref{fig:cycling-path}b, while Figure~\ref{fig:cycling-path}c,~\ref{fig:cycling-path}d,~\ref{fig:cycling-path}e and~\ref{fig:cycling-path}f illustrate images from the four spots. Note that each risk value of the urban spots is the mean risk value of the road section leading to this spot. 

\begin{figure}
\centering
\subfloat[Selected route from the risk map estimated from officially reported accident data~\cite{Castells2019}.]{\includegraphics[width=0.31\textwidth]{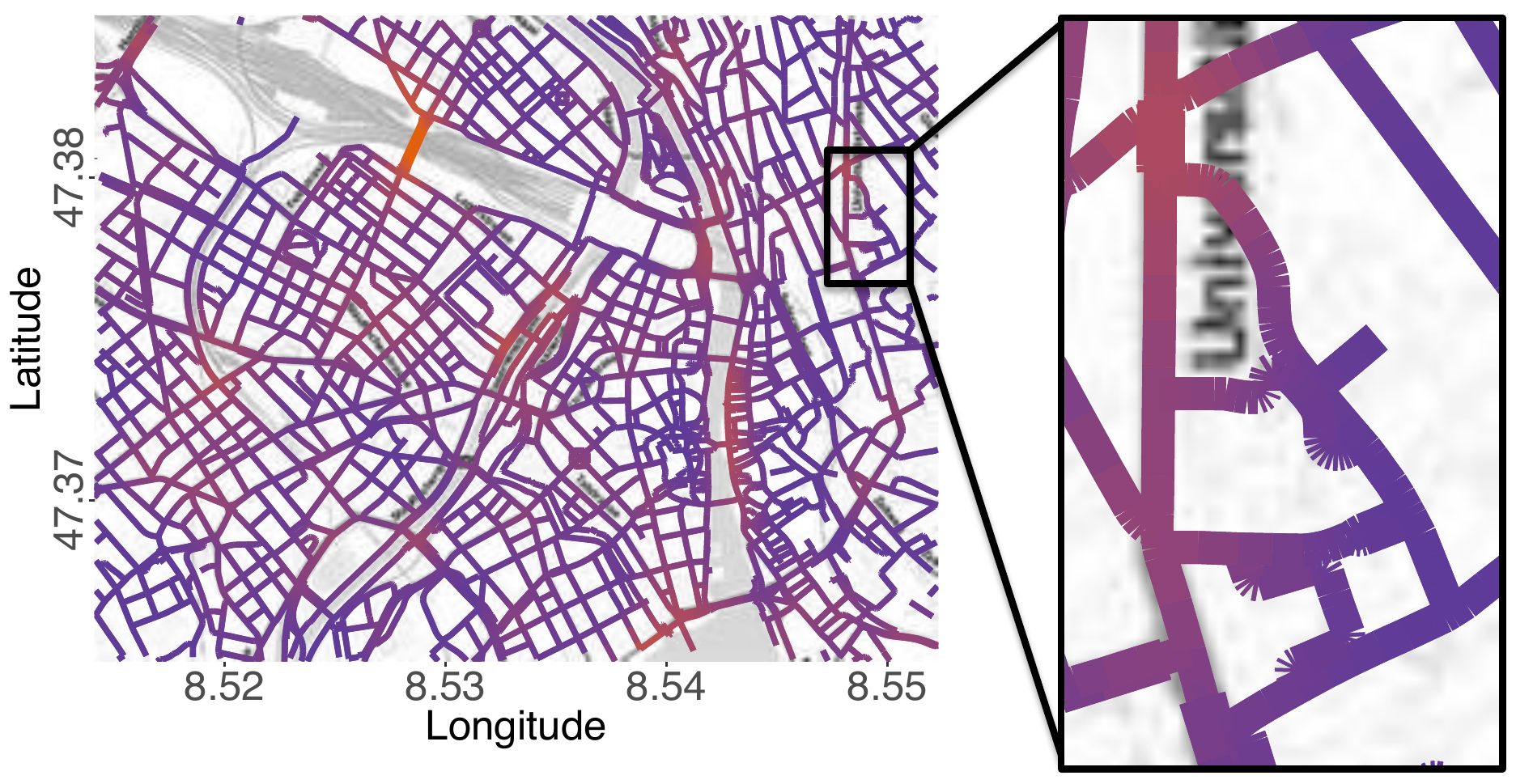}}\hfill
\subfloat[Cycling route and the risk in four urban spots.]{\includegraphics[width=0.158\textwidth]{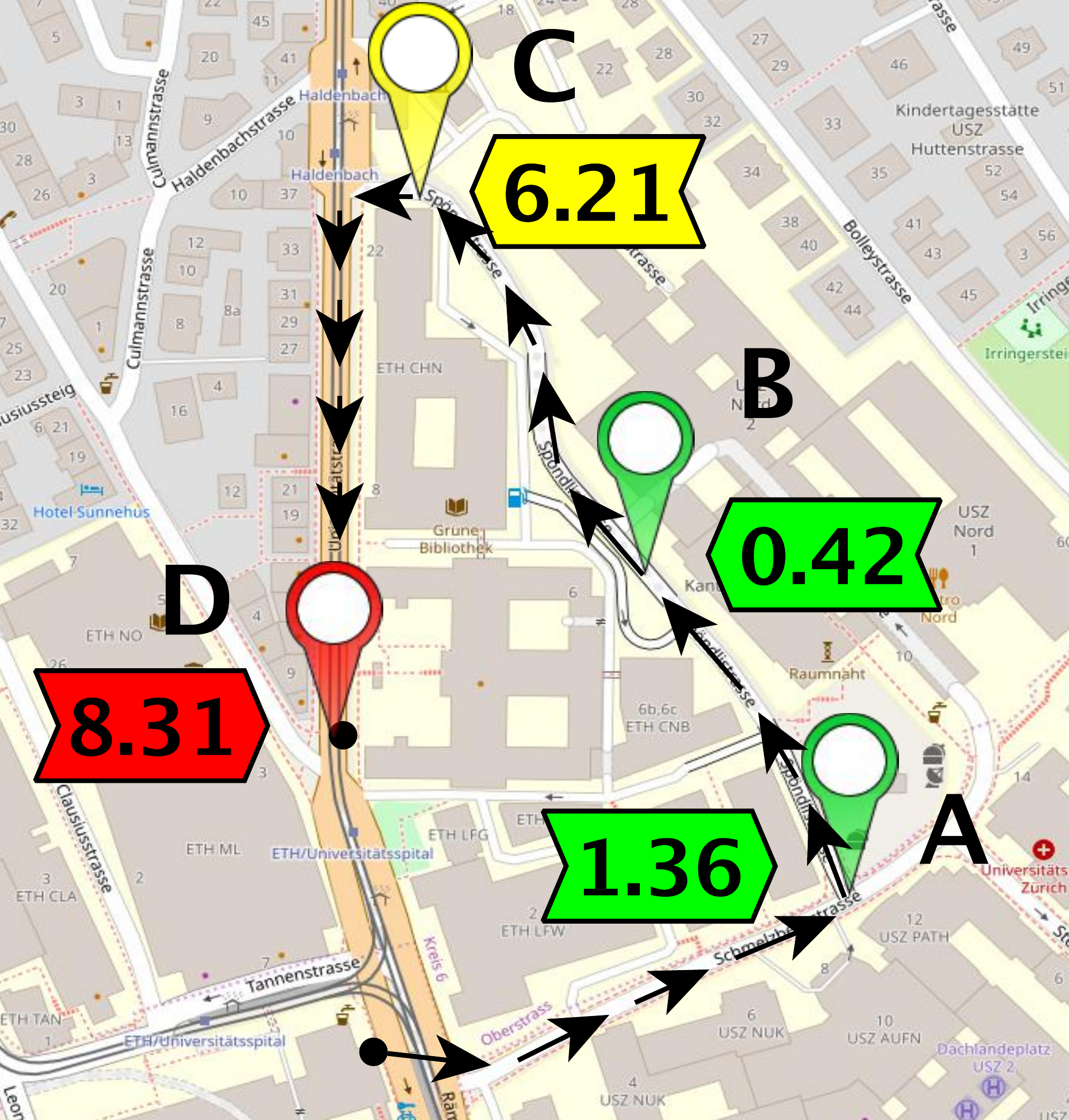}}\\
\subfloat[Spot A with risk value of 1.36.]{\includegraphics[width=0.241\columnwidth]{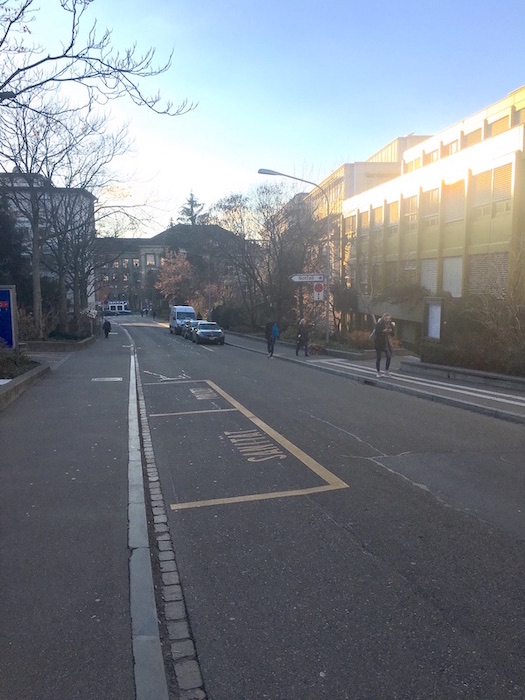}}\hfill
\subfloat[Spot B with risk value of 0.42.]{\includegraphics[width=0.241\columnwidth]{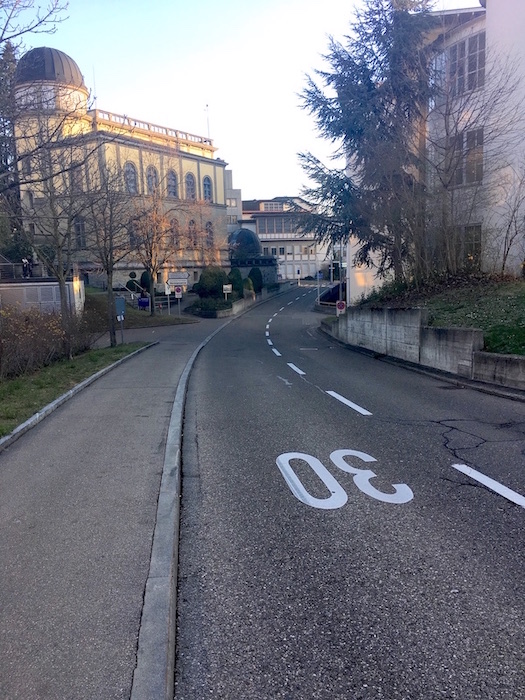}}\hfill
\subfloat[Spot C with risk value of 6.21.]{\includegraphics[width=0.241\columnwidth]{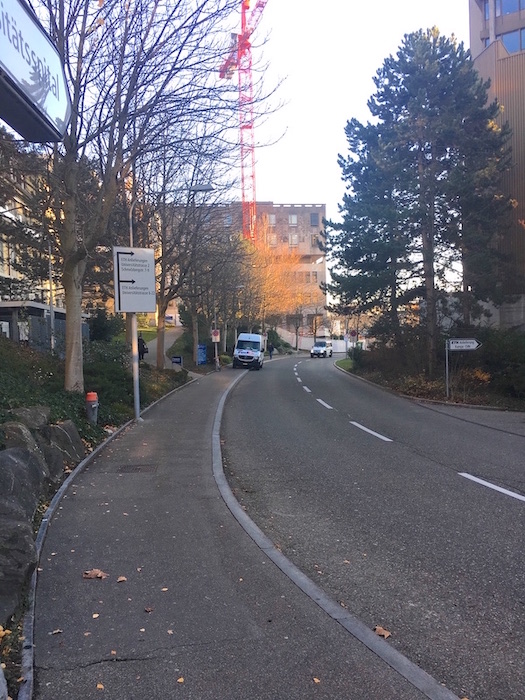}}\hfill
\subfloat[Spot D with risk value of 8.31.]{\includegraphics[width=0.241\columnwidth]{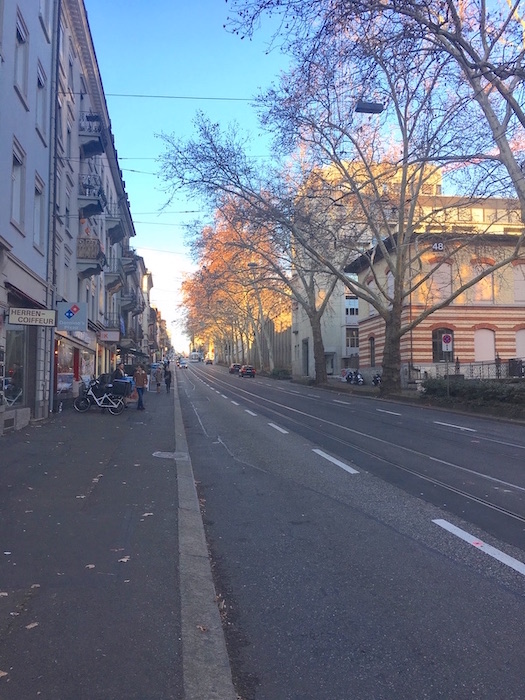}}
\caption{The setup for crowd-sensing cycling safety. The empirical cycling risk values derived from the Federal Roads Office official data of Swiss GeoAdmin~\cite{Castells2019} are compared to the risk values collected by citizens' witness presence.}\label{fig:cycling-path}
\end{figure}

The sequence of the actual cycling risk values across the four urban spots is the baseline for comparison to the perceived cycling risk estimated via the Smart Agora platform. For this purpose, a crowd-sensing asset is designed with Smart Agora using the sequential navigational modality with the same four urban spots of Figure~\ref{fig:cycling-path}b as points of interest. The cycling risk of the road section from the earlier to the next urban spot is assessed when the test cycling user is localized at the next spot, where a likert scale question pops up in the Smart Agora app evaluating cycling risk at a linear scale between 1. \emph{very safe} to 5. \emph{very dangerous}. Answering the questions in all spots completes the cycling trip of a test user and results in a sequence of perceived risk values to compare to the sequence of actual cycling risk values. This comparison is made using both Pearson and Spearman correlation~\cite{Thirumalai2017} for both a numerical and ordinal matching assessment between the two sequences of cycling risk values. Pearson correlation is a measure of linear dependence, i.e. a maximum value of 1 between two sequences of values indicates a perfect linear relationship. However, the actual cycling risk values derived via Gaussian kernel densities~\cite{Castells2019} denote measurements of a non-linear nature. Therefore, the Spearman correlation is used to measure monotonic relationships on the ranking of the cycling risk values. 

Table~\ref{table:cycling-risk} compares the perceived cycling risk values from 11 test users to the actual baseline cycling risk values. All test users cycled over the route on 12.12.2018 around 15:00 with the same provided bike to minimize biases originated from weather, light condition and the condition of different bikes. Correlation values are calculated using the mean and median value of the perceived cycling risk for each urban spot across all users. The Pearson correlation is 0.94 and 0.85 for the mean and median respectively, while the Spearman correlation is 1.0 for both mean and median. 

\begin{table*}
\caption{Perceived cycling risk acquired via the Smart Agora app vs. the actual cycling risk calculated via an empirical model of real-world data~\cite{Castells2019} in the four urban spots of Figure~\ref{fig:cycling-path}. Users' responses are in the range $[1,5]$ with 1 for very safe and 5 for very dangerous.}\label{table:cycling-risk}
\centering
\resizebox{\textwidth}{!}{%
\begin{tabular}{l l l l l l l l l l l l l l l l} 
\toprule
Locations & Test users: & 1 & 2 & 3 & 4 & 5 & 6 & 7 & 8 & 9 & 10 & 11 & Mean & Median & Actual cycling risk~\cite{Castells2019} \\\addlinespace\toprule
Spot A & & 2 & 2 & 2 & 1 & 1 & 1 & 1 & 2 & 2 & 1 & 2 & 1.55 & 2 & 1.36 \\\midrule
Spot B & & 1 & 1 & 1 & 1 & 1 & 1 & 1 & 2 & 1 & 1 & 1 & 1.09 & 1 & 0.42 \\\midrule
Spot C & & 2 & 1 & 1 & 1 & 2 & 3 & 1 & 3 & 4 & 2 & 2 & 2.0 & 2 & 6.21 \\\midrule
Spot D & & 3 & 3 & 3 & 2 & 4 & 4 & 2 & 2 & 3 & 4 & 4 & 3.09 & 3 & 8.31 \\\bottomrule\addlinespace
 & &  &  &  &  &  &  &  &  &  &  & Pearson correlation: & 0.94 & 0.85 &  \\
 & &  &  &  &  &  &  &  &  &  &  & Spearman correlation: & 1.0 & 1.0 & \\\bottomrule
\end{tabular}
}
\end{table*}

Although the number of test users and urban spots is low to reach strong conclusions, the high matching of the two cycling risk estimations in all presented measures suggests that the empirical evidence of cycling accidents matches well with the risk that citizens witness. Therefore, a crowd-based witness presence has a strong potential to verify the status of an urban space and as a result reason about public space more evidently. As an implication, policies designed based on evidence stemming from witness presence promise higher legitimacy for citizens.

\section{Discussion}\label{sec:discussion}

This section discusses dynamic consensus for proving witness presence as well as the role of self-governance and artificial intelligence in the augmented democracy paradigm.

\subsection{Dynamic consensus and self-governance}\label{subsec:}

Proof of witness presence can be validated in a private (permissioned) or public (permissionless) network of nodes running the consensus. For instance, a legally binding deci\-sion-ma\-king process run by city authorities may require a private network of legally representative nodes, similarly to poll clerks in general elections. In case of democratic institutions that may not be well-established, a public network can be a better fit for open self-governed communities encouraging active participation. Moreover, meeting consensus performance requirements using public networks requires access to high-performing public clouds federated by communities or crowd-sourced computational resources deployed by citizens in large-scale. 

An adjustable consensus cost by blockchain platforms~\cite{Tara2019} involves trade-offs between transaction value vs. risk and speed vs. cost. For instance, when performing collective measurements such as the ones in Section~\ref{subsec:testnet}, citizens choices do not all have the same influence on the aggregation accuracy, e.g. the difference from the mean determines the influence. Therefore, witness presence claims can be prioritized based on the influence of citizens' choices on the collective measurements. As a result, accurate estimates are faster with lower transaction costs. Such costs can be further decreased by relaxing the verification rules of the smart contracts executing the proofs of witness presence according to the influence of citizens' choices on the aggregation accuracy. In the application scenario of cycling risk maps (Sectionl~\ref{subsec:cycling-safety}), optimum cycling risk threasholds can be derived to decrease the transaction costs of witness presence (relaxed verification rules) for citizens cycling in risky areas for accidents. 

Such adjustments can be made within community domains that determine validation rules, the number of consensus voters as well as policies/regulations for smart contract execution and data, e.g. General Data Protection Regulation (GDPR). Such domains can also also be used for the self-governance of the augmented democracy paradigm with blockchain providing an efficient and effective automated dispute resolution: reaching consensus on the design of a decision-making process, i.e. navigation modality and collective measurements maps.

\subsection{The role of artificial intelligence}\label{subsec:AI}

Decision support systems such as digital assistants run by artificial intelligence can make decision-making more informed and efficient by overcoming the humans' limitations in congitive bandwidth and the barier of expertise knowledge required to reason about a citizen's choice. However, machine learning algorithms often require sensitive personal data to operate and can be used to nudge citizens and undermine democracy~\cite{Helbing2019}. For instance, the spread of fake news in social media can influence results of elections and therefore massive manipulation of democratic processes is possible using intelligent algorithms~\cite{Allcott2017}. This paper distinguishes two socially responsible and ethically aligned applicability scenarios of artificial intelligence in the proposed augmented democracy paradigm: (i) \emph{local intelligence} and (ii) \emph{collective intelligence}. 

Local intelligence concerns the use of open-source machine learning algorithms that run locally at personal devices of citizens. These algorithms make use of localized or remote open data and they can be used to assist citizens in reaching complex decisions. For instance, a distributed content-based recommender algorithm for more sustainable grocery product choices can make use of public product data related to sustainability. Representation models of these product data can be computed by official authorities and environmental organizations before transferred to citizens' smart phone for personalization~\cite{Hormann2019}. The limitation of local intelligence is that it assists decisions taken from an individual's perspective and it cannot address complex coordination problems that involve several citizens. 

Collective intelligence can address such coordination problems, though the challenge of privacy and transparency remains subject of active research. The concept of \emph{federated learning} is a promising approach for supervised machine learning algorithms and is based on the concept ``bring the code to the data, instead of the data to the code"~\cite{Bonawitz2019,Mcmahan2016}. The concept of \emph{collective learning} is introduced for solving NP hard combinatorial optimization problems in a fully decentralizated fashion given citizens' constraints on privacy and autonomy~\cite{Pournaras2018}. In the augmented democracy paradigm, collective learning can address tragedy of the commons problems in which citizens' choices need to satisfy both individual and collective objectives. Collective learning has been applied\footnote{EPOS, the \emph{Economic Planning and Optimized Selections} is the project studying collective learning~\cite{EPOS2019}.} to application scenarios of sharing economies, e.g. reducing demand power peaks, load-balancing of bike sharing stations, charging control of electric vehicles, traffic flow optimization and other.

%\subsection{Use cases and impact}\label{subsec:use-cases}
%
%%Klinglmayr2017
%
%
%Tax reduction in an area as the means to activate local citizens
%
%take photos to prove presence 
%
%road network, cars verification?
%
%immigrants - citizens leaving tokens  to incentivize their geographical integration! ;)
%
%Humanitarian Aid - We think it is incredibly important to focus on use cases such as humanitarian aid airdrops. While our technology has broad commercial uses, our team is passionate about seeing the project bring new ways to enable ordinary people use cryptocurrency, and an ideal win/win use case is the ability to transfer crypto to people who need assistance. Blockchain and cryptocurrency can often seem intimidating to people, but mass adoption is the direction things are taking. Platin wants to ensure that everyone is able to access and utilize this technology to utilize decentralized funds in a beneficial and potentially life-saving way.
%
%Other cryptocurrencies may be geofenced to the region in distress.

\section{Conclusion and Future Work}\label{sec:conclusion}

This paper concludes that the proposed augmented democracy paradigm is a promising endeavor for building sustainable and participatory Smart Cities. A holistic approach for augmented democracy is introduced based on three pillars that cover participatory crowd-sensing, proof of witness presence and real-time collective measurements. Smart Agora can model a broad spectrum of collective decision-making scenarios given the different types of collected data and navigational modalities. Proving witness presence becomes a cornerstone to a more informed and responsible decision-making. The cycling safety use case scenario illustrated in this paper confirms the accurate information acquired via wisdom of the crowd. Moreover, witness presence has the potential to cultivate high level of engagement and participation integrated in the citizens' daily life and the public space they belong. Linking real-time collective measurements to witness presence provides an added value to crowd-sourced data analytics made by citizens, for citizens. This paper shows how blockchain consensus and crypto-economic design can realize such a grand vision by validating location proofs and incentivizing physical presence. Several localization a\-pproa\-ches are reviewed. An experimental testnet scenario is designed and launched to provide a first technical proof of concept of the proposed augmented democracy paradigm. 

Future work focuses on addressing the limitations of this work. These includes the expansion of the testnet scenario with smart contracts running in the blockchain and providing more advanced and secure proofs of witness presence, beyond GPS and by composing complex social proofs. The influence of mobility patterns and infrastructure on transaction costs and latency requires a further dedicated study. Re\-ly\-ing on to\-ken cu\-ra\-ted registries, for instance the ones of FOAM~\cite{FOAM2019}, for the participation of test users is also subject of future work. Moreover, further use cases in conjunction with city authorities and local communities are required to assess what navigational modalities and collective measurements maps find applicability in real-world. The role of self-governance and an ethically aligned artificial intelligence are expected to play a key role in realizing augmented democracy at large-scale.

\section*{Acknowledgment}

The author would like to thank Prof. Dr. Cesar Hidalgo for his inspiring initiative on augmented democracy as well as for the honor to award to this work the Augmented Democracy Prize (https://www.peopledemocracy.com/prize). Moreover, the author would like to thank Atif Nabi Ghulam for his development contributions in Smart Agora as well as Edward Gaere and Renato Kunz for supporting the development and deployment of the testnet scenario. The author would like to further thank the test users for their participation and feedback. David Castells Graells and Christopher Salahub have especially supported the cycling safety data collection process and contributed to the earlier cycling risk model. Dr. Alexey Gokhberg provided the Hive interfaces to Smart Agora, while earlier author's work together with Jovan Nikolic and all team members of the Nervousnet project put technical foundations for this work. In addition, the author would like to thank his 2018 students of the course ``Data Science in Techno-socio-economic Systems" at ETH Zurich who used Smart Agora and provided invaluable feedback. Many thanks go to author's Empower Polis team members as well as the Institute of Science Technology and Policy (ISTP) of ETH Zurich for running the ETH Policy Challenge and providing a venue to cultivate ideas for digital democracy. Finally, the author would like to heartily thank Prof. Dr. Dirk Helbing and the Chair of Computational Social Science at ETH Zurich for encouraging and supporting this research. 

\appendix

\section{Mobility Traces}\label{sec:traces}

Figure~\ref{fig:GPS-traces} shows the localization traces of the test users for different localization radius. 

\begin{figure*}
\centering	
\subfloat[Group 1, User 1, 50m radius.]{\includegraphics[width=0.31\textwidth]{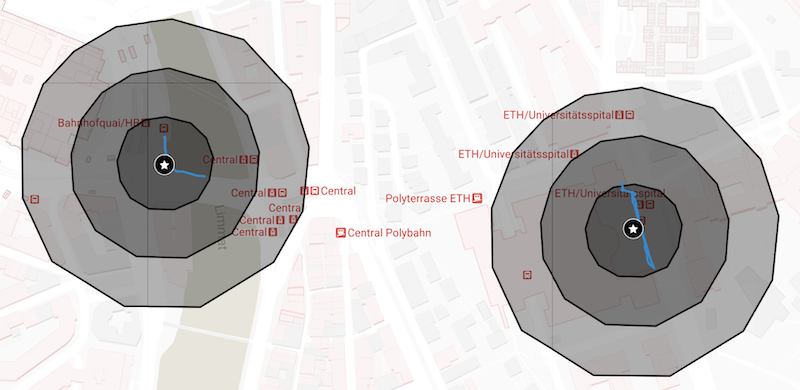}}\label{fig:GRPOUP-01-USER-01-50}\hfill
\subfloat[Group 1, User 2, 100m radius.]{\includegraphics[width=0.31\textwidth]{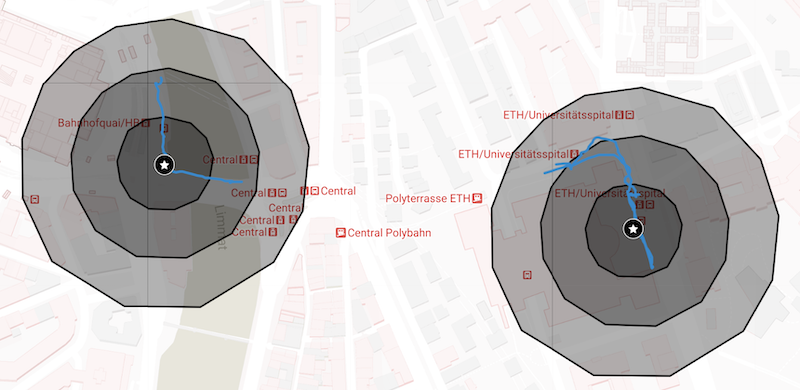}}\label{fig:GRPOUP-01-USER-04-100}\hfill
\subfloat[Group 1, User 3, 150m radius.]{\includegraphics[width=0.31\textwidth]{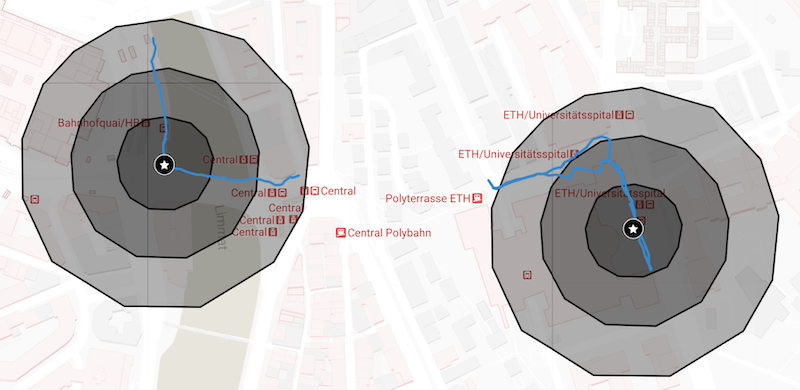}}\label{fig:GRPOUP-01-USER-02-150}\hfill
\subfloat[Group 2, User 4, 50m radius.]{\includegraphics[width=0.31\textwidth]{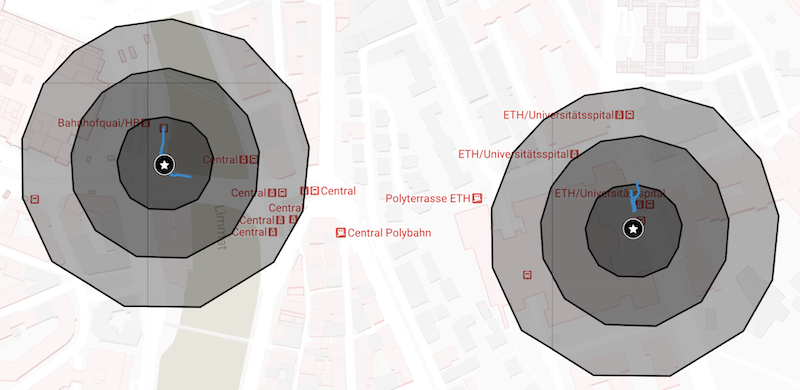}}\label{fig:GRPOUP-02-USER-04-50}\hfill
\subfloat[Group 2, User 5, 100m radius.]{\includegraphics[width=0.31\textwidth]{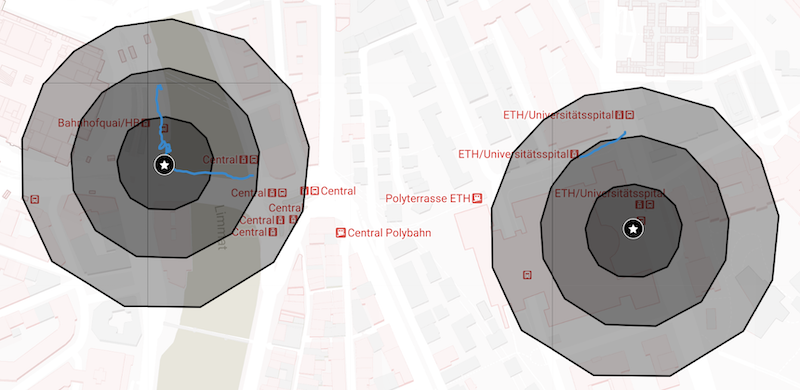}}\label{fig:GRPOUP-02-USER-05-100}\hfill
\subfloat[Group 2, User 6, 150m radius.]{\includegraphics[width=0.31\textwidth]{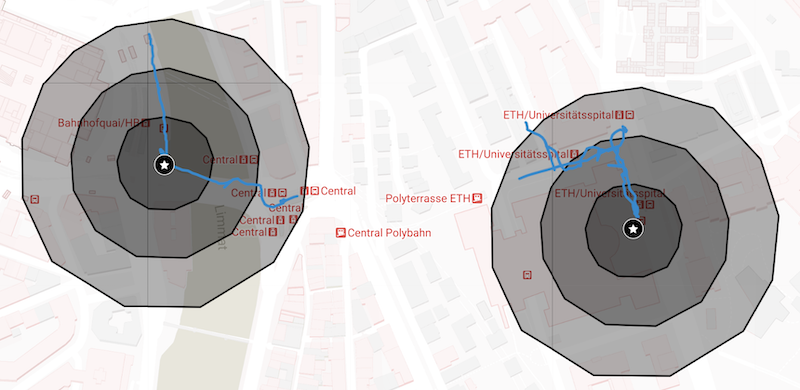}}\label{fig:GRPOUP-02-USER-6-150}\hfill
\caption{GPS traces of test users belonging to different group and having a different localization radius.}\label{fig:GPS-traces}
\end{figure*}


%\bibliographystyle{cas-model2-names}
\bibliographystyle{plainnat}

\bibliography{Smart-Agora} 

\begin{thebibliography}{134}
\providecommand{\natexlab}[1]{#1}
\providecommand{\url}[1]{\texttt{#1}}
\expandafter\ifx\csname urlstyle\endcsname\relax
  \providecommand{\doi}[1]{doi: #1}\else
  \providecommand{\doi}{doi: \begingroup \urlstyle{rm}\Url}\fi

\bibitem[Adh()]{Adhocracy2019}
Adhocracy.
\newblock https://liqd.net.
\newblock (last accessed: April 2020).

\bibitem[Ago()]{Agora2019}
Agora.
\newblock https://www.agora.vote.
\newblock (last accessed: April 2020).

\bibitem[Air()]{Airesis2019}
Airesis.
\newblock https://airesis.eu.
\newblock (last accessed: April 2020).

\bibitem[CON()]{CONSUL2019}
Consul.
\newblock http://consulproject.org.
\newblock (last accessed: April 2020).

\bibitem[Cro({\natexlab{a}})]{Crossiety2019}
Crossiety.
\newblock https://www.crossiety.ch, {\natexlab{a}}.
\newblock (last accessed: April 2020).

\bibitem[Cro({\natexlab{b}})]{CrowdWater2019}
{Crowd Water}.
\newblock https://www.crowdwater.ch/en/welcome-to-crowdwater/, {\natexlab{b}}.
\newblock (last accessed: April 2020).

\bibitem[DIA()]{DIAS2019}
{DIAS: Dynamic Intelligent Aggregation Service}.
\newblock http://dias-net.org.
\newblock (last accessed: April 2020).

\bibitem[Dec()]{Decidim2019}
Decidim.
\newblock https://decidim.org.
\newblock (last accessed: April 2020).

\bibitem[Del()]{Deliberatorium2019}
Deliberatorium.
\newblock http://deliberatorium.mit.edu.
\newblock (last accessed: April 2020).

\bibitem[Dem({\natexlab{a}})]{DemocracyEarth2020}
{Democracy Earth Paper}.
\newblock https://github.com/DemocracyEarth/paper, {\natexlab{a}}.
\newblock (last accessed: April 2020).

\bibitem[Dem({\natexlab{b}})]{DemocracyOS2019}
{DemocracyOS}.
\newblock http://democracyos.org, {\natexlab{b}}.
\newblock (last accessed: April 2020).

\bibitem[Dis()]{Discourse2019}
Discourse.
\newblock https://www.discourse.org.
\newblock (last accessed: April 2020).

\bibitem[EPO()]{EPOS2019}
{EPOS: Economic Planning and Optimized Selections}.
\newblock http://epos-net.org.
\newblock (last accessed: April 2020).

\bibitem[ETH()]{ETHPolicyChallenge2019}
{ETH Policy Challenge}.
\newblock http://www.policychallenge.ch.
\newblock (last accessed: April 2020).

\bibitem[FOA()]{FOAMResearch2019}
{FOAM Research Repository}.
\newblock https://github.com/f-o-a-m/public-research.
\newblock (last accessed: April 2020).

\bibitem[Fol()]{FollowMyVote2019}
{Follow My Vote}.
\newblock https://followmyvote.com.
\newblock (last accessed: April 2020).

\bibitem[Hiv()]{Hive2019}
Hive.
\newblock https://github.com/nytlabs/hive.
\newblock (last accessed: April 2020).

\bibitem[Nov()]{Novoville2019}
Novoville.
\newblock https://novoville.com.
\newblock (last accessed: April 2020).

\bibitem[PWN()]{PWNEDGPSWatshes2019}
Researcher prints 'pwned!' on hundreds of gps watches' maps due to unfixed api.
\newblock
  https://www.zdnet.com/article/researcher-prints-pwned-on-hundreds-of-gps-watches-maps-due-to-unfixed-api/.
\newblock (last accessed: April 2020).

\bibitem[Pla()]{PlacePulse2019}
{Place Pulse}.
\newblock http://pulse.media.mit.edu.
\newblock (last accessed: April 2020).

\bibitem[Pol()]{Polybahn2019}
Polybahn.
\newblock https://en.wikipedia.org/wiki/Polybahn.
\newblock (last accessed: April 2020).

\bibitem[SIK()]{SIKORKA2019}
{SIKORKA - Proof of Presence}.
\newblock http://sikorka.io.
\newblock (last accessed: April 2020).

\bibitem[Scu()]{Scuttlebutt2019}
Scuttlebutt.
\newblock https://www.scuttlebutt.nz.
\newblock (last accessed: April 2020).

\bibitem[Sta()]{StadtZurich2019}
Participation in zurich city.
\newblock https://www.stadt-zuerich.ch/partizipation.
\newblock (last accessed: April 2020).

\bibitem[Swi()]{SwissGeoAdmin2019}
{Swiss GeoAdmin}.
\newblock
  https://map.geo.admin.ch/?topic=vu\&lang=\-de\-\&bg\-Layer=\-\\ch.\-swisstopo.\-pixelkarte-grau\&layers=\-ch.\-astra.\-unfaelle-personenschaeden\_alle\&layers\_timestamp=99990101\&\-catalog\-Nodes\-=\-1318.
\newblock (last accessed: April 2020).

\bibitem[The()]{TheThingsNetwork2019}
T{he Things Network}.
\newblock https://www.thethingsnetwork.org.
\newblock (last accessed: April 2020).

\bibitem[Uni()]{UniquID2020}
{UniquID}.
\newblock https://uniquid.com.
\newblock (last accessed: April 2020).

\bibitem[Vot()]{Votetandem2019}
Votetandem.
\newblock https://votetandem.org.
\newblock (last accessed: April 2020).

\bibitem[WeC()]{WeCollect2019}
{WeCollect}.
\newblock https://wecollect.ch.
\newblock (last accessed: April 2020).

\bibitem[FOA(2018)]{FOAM2019}
{FOAM White Paper}.
\newblock https://www.\-foam.\-space/\-publicAssets\-/FOAM\_Whitepaper.pdf,
  January 2018.
\newblock (last accessed: April 2020).

\bibitem[C4A(2019)]{C4ADS2019}
{Above Us Only Stars - Exposing GPS Spoofing in Russia and Syria}.
\newblock Technical report, Center for Advanced Defense Studies (C4ADS), March
  2019.

\bibitem[Pla(2019)]{Platin2019}
Platin: Proof of location blockchain - white paper.
\newblock https://platin.io/assets/whitepaper/Platin\_Whitepaper\_v3.01.pdf,
  April 2019.
\newblock (last accessed: April 2020).

\bibitem[Abayomi-Zannu et~al.(2019)Abayomi-Zannu, Odun-Ayo, and
  Barka]{Abayomi2019}
TP~Abayomi-Zannu, IA~Odun-Ayo, and TF~Barka.
\newblock A proposed mobile voting framework utilizing blockchain technology
  and multi-factor authentication.
\newblock In \emph{Journal of Physics: Conference Series}, volume 1378, page
  032104. IOP Publishing, 2019.

\bibitem[Agadakos et~al.(2017)Agadakos, Polakis, and
  Portokalidis]{Agadakos2017}
Ioannis Agadakos, Jason Polakis, and Georgios Portokalidis.
\newblock Techu: Open and privacy-preserving crowdsourced gps for the masses.
\newblock In \emph{Proceedings of the 15th Annual International Conference on
  Mobile Systems, Applications, and Services}, pages 475--487. ACM, 2017.

\bibitem[Aitamurto et~al.()Aitamurto, Galli, and Salminen]{Aitamurto2014}
Tanja Aitamurto, Jorge~Saldivar Galli, and Juho Salminen.
\newblock Self-selection in crowdsourced democracy: A bug or a feature.
\newblock In \emph{International Conference on Supporting Group Work (GROUP
  2014. The Morphing Organization: Rethinking Groupwork Systems in the Era of
  Crowdwork.} ACM.

\bibitem[Aladawy et~al.(2018)Aladawy, Beckers, and Pape]{Aladawy2018}
Dina Aladawy, Kristian Beckers, and Sebastian Pape.
\newblock Persuaded: Fighting social engineering attacks with a serious game.
\newblock In \emph{International Conference on Trust and Privacy in Digital
  Business}, pages 103--118. Springer, 2018.

\bibitem[Allcott and Gentzkow(2017)]{Allcott2017}
Hunt Allcott and Matthew Gentzkow.
\newblock Social media and fake news in the 2016 election.
\newblock \emph{Journal of economic perspectives}, 31\penalty0 (2):\penalty0
  211--36, 2017.

\bibitem[Allen et~al.(2017)Allen, Berg, Lane, and Potts]{Allen2017}
Darcy~WE Allen, Chris Berg, Aaron Lane, and Jason Potts.
\newblock The economics of crypto-democracy.
\newblock In \emph{Proceedings of the IJCAI 2017 Workshop on Linked Democracy:
  Artificial Intelligence for Democratic Innovation}, pages 63--73, August
  2017.

\bibitem[America and the Caribbean Regional Office. Sustainable Development
  Unit. Brazil-Country Management~Unit(2008)]{World2008}
World Bank.~Latin America and the Caribbean Regional Office. Sustainable
  Development Unit. Brazil-Country Management~Unit.
\newblock \emph{Brazil: Toward a more inclusive and effective participatory
  budget in Porto Alegre}.
\newblock World Bank, 2008.

\bibitem[{Amoretti} et~al.(2018){Amoretti}, {Brambilla}, {Medioli}, and
  {Zanichelli}]{Amoretti2018}
M.~{Amoretti}, G.~{Brambilla}, F.~{Medioli}, and F.~{Zanichelli}.
\newblock Blockchain-based proof of location.
\newblock In \emph{2018 IEEE International Conference on Software Quality,
  Reliability and Security Companion (QRS-C)}, pages 146--153, July 2018.
\newblock \doi{10.1109/QRS-C.2018.00038}.

\bibitem[Arag{\'o}n et~al.(2017)Arag{\'o}n, Kaltenbrunner, Calleja-L{\'o}pez,
  Pereira, Monterde, Barandiaran, and G{\'o}mez]{Aragon2017}
Pablo Arag{\'o}n, Andreas Kaltenbrunner, Antonio Calleja-L{\'o}pez, Andr{\'e}s
  Pereira, Arnau Monterde, Xabier~E Barandiaran, and Vicen{\c{c}} G{\'o}mez.
\newblock Deliberative platform design: The case study of the online
  discussions in decidim barcelona.
\newblock In \emph{International Conference on Social Informatics}, pages
  277--287. Springer, 2017.

\bibitem[Ballandies et~al.(2018)Ballandies, Dapp, and
  Pournaras]{Ballandies2018}
Mark~C Ballandies, Marcus~M Dapp, and Evangelos Pournaras.
\newblock Decrypting distributed ledger design-taxonomy, classification and
  blockchain community evaluation.
\newblock \emph{arXiv preprint arXiv:1811.03419}, 2018.

\bibitem[Beckers and Pape(2016)]{Beckers2016}
Kristian Beckers and Sebastian Pape.
\newblock A serious game for eliciting social engineering security
  requirements.
\newblock In \emph{2016 IEEE 24th International Requirements Engineering
  Conference (RE)}, pages 16--25. IEEE, 2016.

\bibitem[Berg(2017)]{Perg2017}
Janne Berg.
\newblock Political participation in the form of online petitions: A comparison
  of formal and informal petitioning.
\newblock \emph{International Journal of E-Politics (IJEP)}, 8\penalty0
  (1):\penalty0 14--29, 2017.

\bibitem[Bielenberg et~al.(2012)Bielenberg, Helm, Gentilucci, Stefanescu, and
  Zhang]{Bielenberg2012}
Ames Bielenberg, Lara Helm, Anthony Gentilucci, Dan Stefanescu, and Honggang
  Zhang.
\newblock The growth of diaspora-a decentralized online social network in the
  wild.
\newblock In \emph{2012 Proceedings IEEE INFOCOM Workshops}, pages 13--18.
  IEEE, 2012.

\bibitem[Bonawitz et~al.(2019)Bonawitz, Eichner, Grieskamp, Huba, Ingerman,
  Ivanov, Kiddon, Konecny, Mazzocchi, McMahan, et~al.]{Bonawitz2019}
Keith Bonawitz, Hubert Eichner, Wolfgang Grieskamp, Dzmitry Huba, Alex
  Ingerman, Vladimir Ivanov, Chloe Kiddon, Jakub Konecny, Stefano Mazzocchi,
  H~Brendan McMahan, et~al.
\newblock Towards federated learning at scale: System design.
\newblock \emph{arXiv preprint arXiv:1902.01046}, 2019.

\bibitem[Bornholdt et~al.(2019)Bornholdt, Reher, and Skwarek]{Bornholdt2019}
Lorenz Bornholdt, Julian Reher, and Volker Skwarek.
\newblock Proof-of-location: A method for securing sensor-data-communication in
  a byzantine fault tolerant way.
\newblock In \emph{Mobile Communication-Technologies and Applications; 24.
  ITG-Symposium}, pages 1--6. VDE, 2019.

\bibitem[Bungale and Sridhar(2003)]{Bungale2003}
P~Bungale and Swaroop Sridhar.
\newblock Electronic voting-a survey.
\newblock \emph{Dep. Comput. Sci. Johns Hopkins Univ}, 2003.

\bibitem[Castells-Graells et~al.(2019)Castells-Graells, Salahub, and
  Pournaras]{Castells2019}
David Castells-Graells, Christopher Salahub, and Evangelos Pournaras.
\newblock On cycling risk and discomfort: urban safety mapping and bike route
  recommendations.
\newblock \emph{Computing}, pages 1--16, 2019.

\bibitem[Chen et~al.(2017)Chen, Thombre, J{\"a}rvinen, Lohan, Al{\'e}n-Savikko,
  Lepp{\"a}koski, Bhuiyan, Bu-Pasha, Ferrara, Honkala, et~al.]{Chen2017}
Liang Chen, Sarang Thombre, Kimmo J{\"a}rvinen, Elena~Simona Lohan, Anette
  Al{\'e}n-Savikko, Helena Lepp{\"a}koski, M~Zahidul~H Bhuiyan, Shakila
  Bu-Pasha, Giorgia~Nunzia Ferrara, Salomon Honkala, et~al.
\newblock {Robustness, security and privacy in location-based services for
  future IoT: A survey}.
\newblock \emph{IEEE Access}, 5:\penalty0 8956--8977, 2017.

\bibitem[Condorcet and De(1793)]{Condorcet1793}
Jean-Antoine-Nicolas de~Caritat Condorcet and Marquis De.
\newblock Plan de constitution pr{\'e}sent{\'e} {\`a} la convention nationale.
\newblock \emph{Paris: Imprimerie Nationale}, 1793.

\bibitem[Contucci et~al.(2016)Contucci, Panizzi, Ricci-Tersenghi, and
  S{\^\i}rbu]{Contucci2016}
Pierluigi Contucci, Emanuele Panizzi, Federico Ricci-Tersenghi, and Alina
  S{\^\i}rbu.
\newblock Egalitarianism in the rank aggregation problem: a new dimension for
  democracy.
\newblock \emph{Quality \& Quantity}, 50\penalty0 (3):\penalty0 1185--1200,
  2016.

\bibitem[Danaf et~al.(2019)Danaf, Atasoy, De~Azevedo, Ding-Mastera, Abou-Zeid,
  Cox, Zhao, and Ben-Akiva]{Danaf2019}
Mazen Danaf, Bilge Atasoy, Carlos~Lima De~Azevedo, Jing Ding-Mastera, Maya
  Abou-Zeid, Nathaniel Cox, Fang Zhao, and Moshe Ben-Akiva.
\newblock Context-aware stated preferences with smartphone-based travel
  surveys.
\newblock \emph{Journal of Choice Modelling}, 2019.

\bibitem[Dasu et~al.(2018)Dasu, Kanza, and Srivastava]{Dasu2018}
Tamraparni Dasu, Yaron Kanza, and Divesh Srivastava.
\newblock Unchain your blockchain.
\newblock In \emph{Proc. Symposium on Foundations and Applications of
  Blockchain}, volume~1, pages 16--23, 2018.

\bibitem[Dubey et~al.(2016)Dubey, Naik, Parikh, Raskar, and Hidalgo]{Dubey2016}
Abhimanyu Dubey, Nikhil Naik, Devi Parikh, Ramesh Raskar, and C{\'e}sar~A
  Hidalgo.
\newblock Deep learning the city: Quantifying urban perception at a global
  scale.
\newblock In \emph{European conference on computer vision}, pages 196--212.
  Springer, 2016.

\bibitem[Emerson(2020)]{Emerson2020}
Peter Emerson.
\newblock Majoritarian democracy: The catalyst of populism.
\newblock In \emph{Majority Voting as a Catalyst of Populism}, pages 185--189.
  Springer, 2020.

\bibitem[Etter et~al.(2018)Etter, Colleoni, Illia, Meggiorin, and
  D’Eugenio]{Etter2018}
Michael Etter, Elanor Colleoni, Laura Illia, Katia Meggiorin, and Antonino
  D’Eugenio.
\newblock Measuring organizational legitimacy in social media: Assessing
  citizens’ judgments with sentiment analysis.
\newblock \emph{Business \& Society}, 57\penalty0 (1):\penalty0 60--97, 2018.

\bibitem[Falk and Tsoukalas(2018)]{Falk2018}
Brett~Hemenway Falk and Gerry Tsoukalas.
\newblock Token weighted crowdsourcing.
\newblock Technical report, Working Paper, 2018.

\bibitem[Fisher and Shorrocks(2018)]{Fisher2018}
Stephen~D Fisher and Rosalind Shorrocks.
\newblock Collective failure? lessons from combining forecasts for the uk's
  referendum on eu membership.
\newblock \emph{Journal of Elections, Public Opinion and Parties}, 28\penalty0
  (1):\penalty0 59--77, 2018.

\bibitem[Fishkin(2011)]{Fishkin2011}
James~S Fishkin.
\newblock Deliberative democracy and constitutions.
\newblock \emph{Social philosophy and policy}, 28\penalty0 (1):\penalty0
  242--260, 2011.

\bibitem[Fung(2015)]{Fung2015}
Archon Fung.
\newblock Putting the public back into governance: The challenges of citizen
  participation and its future.
\newblock \emph{Public Administration Review}, 75\penalty0 (4):\penalty0
  513--522, 2015.

\bibitem[Gibson et~al.(2016)Gibson, Krimmer, Teague, and Pomares]{Gibson2016}
J~Paul Gibson, Robert Krimmer, Vanessa Teague, and Julia Pomares.
\newblock A review of e-voting: the past, present and future.
\newblock \emph{Annals of Telecommunications}, 71\penalty0 (7-8):\penalty0
  279--286, 2016.

\bibitem[Gr{\v{c}}ar et~al.(2017)Gr{\v{c}}ar, Cherepnalkoski, Mozeti{\v{c}},
  and Novak]{Grvcar2017}
Miha Gr{\v{c}}ar, Darko Cherepnalkoski, Igor Mozeti{\v{c}}, and Petra~Kralj
  Novak.
\newblock Stance and influence of twitter users regarding the brexit
  referendum.
\newblock \emph{Computational social networks}, 4\penalty0 (1):\penalty0 6,
  2017.

\bibitem[Griego et~al.(2017)Griego, Buff, Hayoz, Moise, and
  Pournaras]{Griego2017}
Danielle Griego, Varin Buff, Eric Hayoz, Izabela Moise, and Evangelos
  Pournaras.
\newblock Sensing and mining urban qualities in smart cities.
\newblock In \emph{2017 IEEE 31st International Conference on Advanced
  Information Networking and Applications (AINA)}, pages 1004--1011. IEEE,
  2017.

\bibitem[Haggerty and Samatas(2010)]{Haggerty2010}
Kevin~D Haggerty and Minas Samatas.
\newblock \emph{Surveillance and democracy}.
\newblock Routledge, 2010.

\bibitem[Helbing et~al.(2019)Helbing, Frey, Gigerenzer, Hafen, Hagner,
  Hofstetter, Van Den~Hoven, Zicari, and Zwitter]{Helbing2019}
Dirk Helbing, Bruno~S Frey, Gerd Gigerenzer, Ernst Hafen, Michael Hagner,
  Yvonne Hofstetter, Jeroen Van Den~Hoven, Roberto~V Zicari, and Andrej
  Zwitter.
\newblock Will democracy survive big data and artificial intelligence?
\newblock In \emph{Towards Digital Enlightenment}, pages 73--98. Springer,
  2019.

\bibitem[Hormann et~al.(2019)Hormann, Putz, Rudic, Kastl, Klinglmayr, and
  Pournaras]{Hormann2019}
Leander~B Hormann, Veronika Putz, Branislav Rudic, Christian Kastl, Johannes
  Klinglmayr, and Evangelos Pournaras.
\newblock Augmented shopping experience for sustainable consumption using the
  internet of thing.
\newblock \emph{IEEE Internet of Things Magazine}, 2\penalty0 (3):\penalty0
  46--51, 2019.

\bibitem[Iandoli et~al.(2018)Iandoli, Quinto, Spada, Klein, and
  Calabretta]{Iandoli2018}
Luca Iandoli, Ivana Quinto, Paolo Spada, Mark Klein, and Raffaele Calabretta.
\newblock Supporting argumentation in online political debate: Evidence from an
  experiment of collective deliberation.
\newblock \emph{New Media \& Society}, 20\penalty0 (4):\penalty0 1320--1341,
  2018.

\bibitem[Javali et~al.(2016)Javali, Revadigar, Rasmussen, Hu, and
  Jha]{Javali2016}
Chitra Javali, Girish Revadigar, Kasper~B Rasmussen, Wen Hu, and Sanjay Jha.
\newblock I am alice, i was in wonderland: secure location proof generation and
  verification protocol.
\newblock In \emph{2016 IEEE 41st conference on local computer networks (LCN)},
  pages 477--485. IEEE, 2016.

\bibitem[Jiang et~al.(2020)Jiang, Cao, Krishnamachari, Zhou, and
  Niu]{Jiang2020}
Zhiyuan Jiang, Zixu Cao, Bhaskar Krishnamachari, Sheng Zhou, and Zhisheng Niu.
\newblock {SENATE: A Permissionless Byzantine Consensus Protocol in Wireless
  Networks for Real-Time Internet-of-Things Applications}.
\newblock \emph{IEEE Internet of Things Journal}, 2020.

\bibitem[Khan et~al.(2014)Khan, Zawoad, Haque, and Hasan]{Khan2014}
Rasib Khan, Shams Zawoad, Md~Munirul Haque, and Ragib Hasan.
\newblock ‘who, when, and where?’location proof assertion for mobile
  devices.
\newblock In \emph{IFIP Annual Conference on Data and Applications Security and
  Privacy}, pages 146--162. Springer, 2014.

\bibitem[Khanchandani and Lenzen(2019)]{Khanchandani2019}
Pankaj Khanchandani and Christoph Lenzen.
\newblock Self-stabilizing byzantine clock synchronization with optimal
  precision.
\newblock \emph{Theory of Computing Systems}, 63\penalty0 (2):\penalty0
  261--305, 2019.

\bibitem[Kuperberg et~al.(2019)Kuperberg, Kemper, and Durak]{Kuperberg2019}
Michael Kuperberg, Sebastian Kemper, and Cemil Durak.
\newblock Blockchain usage for government-issued electronic ids: A survey.
\newblock In \emph{International Conference on Advanced Information Systems
  Engineering}, pages 155--167. Springer, 2019.

\bibitem[Lalley and Weyl(2018)]{Lalley2018}
Steven~P Lalley and E~Glen Weyl.
\newblock Quadratic voting: How mechanism design can radicalize democracy.
\newblock In \emph{AEA Papers and Proceedings}, volume 108, pages 33--37, 2018.

\bibitem[Lam et~al.(2015)Lam, Chen, Whittle, Binner, and
  Lawlor-Wright]{Lam2015}
Busayawan Lam, Yu~Ping Chen, Jon Whittle, Jane Binner, and Therese
  Lawlor-Wright.
\newblock Better service design for greater civic engagement.
\newblock \emph{The Design Journal}, 18\penalty0 (1):\penalty0 31--55, 2015.

\bibitem[Lao et~al.(2020)Lao, Li, Hou, Xiao, Guo, and Yang]{Lao2020}
Laphou Lao, Zecheng Li, Songlin Hou, Bin Xiao, Songtao Guo, and Yuanyuan Yang.
\newblock {A Survey of IoT Applications in Blockchain Systems: Architecture,
  Consensus, and Traffic Modeling}.
\newblock \emph{ACM Computing Surveys (CSUR)}, 53\penalty0 (1):\penalty0 1--32,
  2020.

\bibitem[Lazer et~al.(2018)Lazer, Baum, Benkler, Berinsky, Greenhill, Menczer,
  Metzger, Nyhan, Pennycook, Rothschild, et~al.]{Lazer2018}
David~MJ Lazer, Matthew~A Baum, Yochai Benkler, Adam~J Berinsky, Kelly~M
  Greenhill, Filippo Menczer, Miriam~J Metzger, Brendan Nyhan, Gordon
  Pennycook, David Rothschild, et~al.
\newblock The science of fake news.
\newblock \emph{Science}, 359\penalty0 (6380):\penalty0 1094--1096, 2018.

\bibitem[LeDuc(2015)]{Leduc2015}
Lawrence LeDuc.
\newblock Referendums and deliberative democracy.
\newblock \emph{Electoral studies}, 38:\penalty0 139--148, 2015.

\bibitem[Levi et~al.(2009)Levi, Sacks, and Tyler]{Levi2009}
Margaret Levi, Audrey Sacks, and Tom Tyler.
\newblock Conceptualizing legitimacy, measuring legitimating beliefs.
\newblock \emph{American behavioral scientist}, 53\penalty0 (3):\penalty0
  354--375, 2009.

\bibitem[Li et~al.(2016)Li, Zhou, Zhu, and Sun]{Li2016}
Yi~Li, Lu~Zhou, Haojin Zhu, and Limin Sun.
\newblock Privacy-preserving location proof for securing large-scale
  database-driven cognitive radio networks.
\newblock \emph{IEEE Internet of Things Journal}, 3\penalty0 (4):\penalty0
  563--571, 2016.

\bibitem[Lim et~al.(2018)Lim, Fotsing, Almasri, Musa, Kiah, Ang, and
  Ismail]{Lim2018}
Shu~Yun Lim, Pascal~Tankam Fotsing, Abdullah Almasri, Omar Musa, Miss Laiha~Mat
  Kiah, Tan~Fong Ang, and Reza Ismail.
\newblock Blockchain technology the identity management and authentication
  service disruptor: a survey.
\newblock \emph{International Journal on Advanced Science, Engineering and
  Information Technology}, 8\penalty0 (4-2):\penalty0 1735, 2018.

\bibitem[Lo et~al.(2020)Lo, Xu, Staples, and Yao]{Lo2020}
Sin~Kuang Lo, Xiwei Xu, Mark Staples, and Lina Yao.
\newblock Reliability analysis for blockchain oracles.
\newblock \emph{Computers \& Electrical Engineering}, 83:\penalty0 106582,
  2020.

\bibitem[Lyu et~al.(2015)Lyu, Pande, Wang, Zhu, Gu, and Mohapatra]{Lyu2015}
Chen Lyu, Amit Pande, Xinlei Wang, Jindan Zhu, Dawu Gu, and Prasant Mohapatra.
\newblock Clip: Continuous location integrity and provenance for mobile phones.
\newblock In \emph{2015 IEEE 12th International Conference on Mobile Ad Hoc and
  Sensor Systems}, pages 172--180. IEEE, 2015.

\bibitem[Ma et~al.(2019)Ma, Kaneko, Sharma, and Sakurai]{Ma2019}
Limao Ma, Kosuke Kaneko, Subodh Sharma, and Kouichi Sakurai.
\newblock Reliable decentralized oracle with mechanisms for verification and
  disputation.
\newblock In \emph{2019 Seventh International Symposium on Computing and
  Networking Workshops (CANDARW)}, pages 346--352. IEEE, 2019.

\bibitem[Malekpour(2015)]{Malekpour2015}
Mahyar~R Malekpour.
\newblock A self-stabilizing hybrid fault-tolerant synchronization protocol.
\newblock In \emph{2015 IEEE Aerospace Conference}, pages 1--11. IEEE, 2015.

\bibitem[Malekpour(2017)]{Malekpour2017}
Mahyar~R Malekpour.
\newblock An autonomous distributed fault-tolerant local positioning system,
  2017.

\bibitem[McMahan et~al.(2016)McMahan, Moore, Ramage, Hampson,
  et~al.]{Mcmahan2016}
H~Brendan McMahan, Eider Moore, Daniel Ramage, Seth Hampson, et~al.
\newblock Communication-efficient learning of deep networks from decentralized
  data.
\newblock \emph{arXiv preprint arXiv:1602.05629}, 2016.

\bibitem[Musciotto et~al.(2016)Musciotto, Delpriori, Castagno, and
  Pournaras]{Musciotto2016}
Federico Musciotto, Saverio Delpriori, Paolo Castagno, and Evangelos Pournaras.
\newblock Mining social interactions in privacy-preserving temporal networks.
\newblock In \emph{Proceedings of the 2016 IEEE/ACM International Conference on
  Advances in Social Networks Analysis and Mining}, pages 1103--1110. IEEE
  Press, 2016.

\bibitem[Naik et~al.(2014)Naik, Philipoom, Raskar, and Hidalgo]{Naik2014}
Nikhil Naik, Jade Philipoom, Ramesh Raskar, and C{\'e}sar Hidalgo.
\newblock Streetscore-predicting the perceived safety of one million
  streetscapes.
\newblock In \emph{Proceedings of the IEEE Conference on Computer Vision and
  Pattern Recognition Workshops}, pages 779--785, 2014.

\bibitem[Nasrulin et~al.(2018)Nasrulin, Muzammal, and Qu]{Nasrulin2018}
Bulat Nasrulin, Muhammad Muzammal, and Qiang Qu.
\newblock A robust spatio-temporal verification protocol for blockchain.
\newblock In \emph{International Conference on Web Information Systems
  Engineering}, pages 52--67. Springer, 2018.

\bibitem[Nevejan(2009)]{Nevejan2009}
Caroline Nevejan.
\newblock Witnessed presence and the yutpa framework.
\newblock \emph{PsychNology Journal}, 7\penalty0 (1), 2009.

\bibitem[Nieto et~al.(2018)Nieto, Rios, and Lopez]{Nieto2018}
Ana Nieto, Ruben Rios, and Javier Lopez.
\newblock I{oT-Forensics meets privacy: towards cooperative digital
  investigations}.
\newblock \emph{Sensors}, 18\penalty0 (2):\penalty0 492, 2018.

\bibitem[Nissen et~al.(2018)Nissen, Pschetz, Murray-Rust, Mehrpouya,
  Oosthuizen, and Speed]{Nissen2018}
Bettina Nissen, Larissa Pschetz, Dave Murray-Rust, Hadi Mehrpouya, Shaune
  Oosthuizen, and Chris Speed.
\newblock Geocoin: Supporting ideation and collaborative design with smart
  contracts.
\newblock In \emph{Proceedings of the 2018 CHI Conference on Human Factors in
  Computing Systems}, page 163. ACM, 2018.

\bibitem[Norta et~al.(2019)Norta, Matulev{\u\i}cius, and Leiding]{Norta2019}
Alex Norta, Raimundas Matulev{\u\i}cius, and Benjamin Leiding.
\newblock Safeguarding a formalized blockchain-enabled identity-authentication
  protocol by applying security risk-oriented patterns.
\newblock \emph{Computers \& Security}, 2019.

\bibitem[Ober(2008)]{Ober2008}
Josiah Ober.
\newblock \emph{Democracy and knowledge: Innovation and learning in classical
  Athens}.
\newblock Princeton University Press, 2008.

\bibitem[Otte et~al.(2017)Otte, de~Vos, and Pouwelse]{Otte2017}
Pim Otte, Martijn de~Vos, and Johan Pouwelse.
\newblock Trustchain: A sybil-resistant scalable blockchain.
\newblock \emph{Future Generation Computer Systems}, 2017.

\bibitem[Pe{\~n}a-L{\'o}pez(2017)]{Pena2017}
Ismael Pe{\~n}a-L{\'o}pez.
\newblock Citizen participation and the rise of the open source city in spain.
\newblock 2017.

\bibitem[Poblet and Plaza(2017)]{Poblet2017}
Marta Poblet and Enric Plaza.
\newblock Democracy models and civic technologies: Tensions, trilemmas, and
  trade-offs.
\newblock In \emph{Proceedings of the IJCAI 2017 Workshop on Linked Democracy:
  Artificial Intelligence for Democratic Innovation}, pages 51--62, August
  2017.

\bibitem[Poon and Buterin(2017)]{Poon2017}
Joseph Poon and Vitalik Buterin.
\newblock Plasma: Scalable autonomous smart contracts.
\newblock \emph{White paper}, pages 1--47, 2017.

\bibitem[Pournaras and Nikoli{\'c}(2017{\natexlab{a}})]{Pournaras2017e}
Evangelos Pournaras and Jovan Nikoli{\'c}.
\newblock Self-corrective dynamic networks via decentralized reverse
  computations.
\newblock In \emph{2017 IEEE International Conference on Autonomic Computing
  (ICAC)}, pages 11--20. IEEE, 2017{\natexlab{a}}.

\bibitem[Pournaras and Nikoli{\'c}(2017{\natexlab{b}})]{Pournaras2017f}
Evangelos Pournaras and Jovan Nikoli{\'c}.
\newblock On-demand self-adaptive data analytics in large-scale decentralized
  networks.
\newblock In \emph{2017 IEEE 16th International Symposium on Network Computing
  and Applications (NCA)}, pages 1--10. IEEE, 2017{\natexlab{b}}.

\bibitem[Pournaras et~al.(2017)Pournaras, Nikolic, Omerzel, and
  Helbing]{Pournaras2017c}
Evangelos Pournaras, Jovan Nikolic, Alex Omerzel, and Dirk Helbing.
\newblock Engineering democratization in internet of things data analytics.
\newblock In \emph{2017 IEEE 31st International Conference on Advanced
  Information Networking and Applications (AINA)}, pages 994--1003. IEEE, 2017.

\bibitem[Pournaras et~al.(2018)Pournaras, Pilgerstorfer, and
  Asikis]{Pournaras2018}
Evangelos Pournaras, Peter Pilgerstorfer, and Thomas Asikis.
\newblock Decentralized collective learning for self-managed sharing economies.
\newblock \emph{ACM Transactions on Autonomous and Adaptive Systems (TAAS)},
  13\penalty0 (2):\penalty0 10, 2018.

\bibitem[Pournaras et~al.(2019)Pournaras, Gaere, Kunz, and
  Ghulam]{Pournaras2019e}
Evangelos Pournaras, Edward Gaere, Renato Kunz, and Atif~Nabi Ghulam.
\newblock Democratizing data analytics: Crowd-sourcing decentralized collective
  measurements.
\newblock In \emph{13th International Conference on Self-adaptive and
  Self-organizing Systems (SASO 2019)}. IEEE, 2019.

\bibitem[Prelipcean et~al.(2018)Prelipcean, Gid{\'o}falvi, and
  Susilo]{Prelipcean2018}
Adrian~C Prelipcean, Gy{\H{o}}z{\H{o}} Gid{\'o}falvi, and Yusak~O Susilo.
\newblock Meili: A travel diary collection, annotation and automation system.
\newblock \emph{Computers, Environment and Urban Systems}, 70:\penalty0 24--34,
  2018.

\bibitem[Qi et~al.(2017)Qi, Feng, Liu, and Mrad]{Qi2017}
Renming Qi, Chen Feng, Zheng Liu, and Nezih Mrad.
\newblock Blockchain-powered internet of things, e-governance and e-democracy.
\newblock In \emph{E-Democracy for Smart Cities}, pages 509--520. Springer,
  2017.

\bibitem[Ramachandran et~al.(2018)Ramachandran, Radhakrishnan, and
  Krishnamachari]{Ramachandran2018}
Gowri~Sankar Ramachandran, Rahul Radhakrishnan, and Bhaskar Krishnamachari.
\newblock Towards a decentralized data marketplace for smart cities.
\newblock In \emph{2018 IEEE International Smart Cities Conference (ISC2)},
  pages 1--8. IEEE, 2018.

\bibitem[Raza et~al.(2017)Raza, Kulkarni, and Sooriyabandara]{Raza2017}
Usman Raza, Parag Kulkarni, and Mahesh Sooriyabandara.
\newblock Low power wide area networks: An overview.
\newblock \emph{IEEE Communications Surveys \& Tutorials}, 19\penalty0
  (2):\penalty0 855--873, 2017.

\bibitem[Rouillard(2008)]{Rouillard2008}
Jos{\'e} Rouillard.
\newblock Contextual qr codes.
\newblock In \emph{2008 The Third International Multi-Conference on Computing
  in the Global Information Technology (iccgi 2008)}, pages 50--55. IEEE, 2008.

\bibitem[{\c{S}}ahan et~al.(2019){\c{S}}ahan, Ekici, and Bahtiyar]{Csahan2019}
Sercan {\c{S}}ahan, Adil~Furkan Ekici, and {\c{S}}erif Bahtiyar.
\newblock A multi-factor authentication framework for secure access to
  blockchain.
\newblock In \emph{Proceedings of the 2019 5th International Conference on
  Computer and Technology Applications}, pages 160--164, 2019.

\bibitem[Salesses et~al.(2013)Salesses, Schechtner, and Hidalgo]{Salesses2013}
Philip Salesses, Katja Schechtner, and C{\'e}sar~A Hidalgo.
\newblock The collaborative image of the city: mapping the inequality of urban
  perception.
\newblock \emph{PloS one}, 8\penalty0 (7):\penalty0 e68400, 2013.

\bibitem[Schaab et~al.(2017)Schaab, Beckers, and Pape]{Schaab2017}
Peter Schaab, Kristian Beckers, and Sebastian Pape.
\newblock Social engineering defence mechanisms and counteracting training
  strategies.
\newblock \emph{Information \& Computer Security}, 25\penalty0 (2):\penalty0
  206--222, 2017.

\bibitem[Schulze(2011)]{Schulze2011}
Markus Schulze.
\newblock A new monotonic, clone-independent, reversal symmetric, and
  condorcet-consistent single-winner election method.
\newblock \emph{Social Choice and Welfare}, 36\penalty0 (2):\penalty0 267--303,
  2011.

\bibitem[Seibert et~al.(2019)Seibert, Strobl, Etter, Hummer, and van
  Meerveld]{Seibert2019}
Jan Seibert, Barbara Strobl, Simon Etter, Philipp Hummer, and HJ~van Meerveld.
\newblock Virtual staff gauges for crowd-based stream level observations.
\newblock \emph{Frontiers in Earth Science}, 7:\penalty0 70, 2019.

\bibitem[Susskind(2017)]{Susskind2017}
Jane Susskind.
\newblock Decrypting democracy: Incentivizing blockchain voting technology for
  an improved election system.
\newblock \emph{San Diego L. Rev.}, 54:\penalty0 785, 2017.

\bibitem[Taher et~al.(2019)Taher, Nahar, and Hossain]{Taher2019}
Kazi~Abu Taher, Tahmin Nahar, and Syed~Akhter Hossain.
\newblock Enhanced cryptocurrency security by time-based token multi-factor
  authentication algorithm.
\newblock In \emph{2019 International Conference on Robotics, Electrical and
  Signal Processing Techniques (ICREST)}, pages 308--312. IEEE, 2019.

\bibitem[Tara et~al.(2019)Tara, Ivkushkin, Butean, and Turesson]{Tara2019}
Andrei Tara, Kirill Ivkushkin, Alexandru Butean, and Hjalmar Turesson.
\newblock The evolution of blockchain virtual machine architecture towards an
  enterprise usage perspective.
\newblock In \emph{Computer Science On-line Conference}, pages 370--379.
  Springer, 2019.

\bibitem[Tarr et~al.(2019)Tarr, Lavoie, Meyer, and Tschudin]{Tarr2019}
Dominic Tarr, Erick Lavoie, Aljoscha Meyer, and Christian Tschudin.
\newblock Secure scuttlebutt: An identity-centric protocol for subjective and
  decentralized applications.
\newblock In \emph{Proceedings of the 6th ACM Conference on Information-Centric
  Networking}, pages 1--11. ACM, 2019.

\bibitem[Thirumalai et~al.(2017)Thirumalai, Chandhini, and
  Vaishnavi]{Thirumalai2017}
Chandrasegar Thirumalai, Swapna~Anupriya Chandhini, and M~Vaishnavi.
\newblock Analysing the concrete compressive strength using pearson and
  spearman.
\newblock In \emph{2017 International conference of Electronics, Communication
  and Aerospace Technology (ICECA)}, volume~2, pages 215--218. IEEE, 2017.

\bibitem[Tippenhauer et~al.(2011)Tippenhauer, P{\"o}pper, Rasmussen, and
  Capkun]{Tippenhauer2011}
Nils~Ole Tippenhauer, Christina P{\"o}pper, Kasper~Bonne Rasmussen, and Srdjan
  Capkun.
\newblock On the requirements for successful gps spoofing attacks.
\newblock In \emph{Proceedings of the 18th ACM conference on Computer and
  communications security}, pages 75--86. ACM, 2011.

\bibitem[Ugwu et~al.(2018)Ugwu, Okpala, Oham, and Nwakanma]{Ugwu2018}
Marcel~C Ugwu, Izunna~U Okpala, Collins~I Oham, and Cosmas~I Nwakanma.
\newblock A tiered blockchain framework for vehicular forensics.
\newblock \emph{International Journal of Network Security \& Its Applications
  (IJNSA) Vol}, 10, 2018.

\bibitem[Urbinati(2004)]{Urbinati2004}
Nadia Urbinati.
\newblock Condorcet’s democratic theory of representative government.
\newblock \emph{European journal of political theory}, 3\penalty0 (1):\penalty0
  53--75, 2004.

\bibitem[Urbinati and Warren(2008)]{Urbinati2008}
Nadia Urbinati and Mark~E Warren.
\newblock The concept of representation in contemporary democratic theory.
\newblock \emph{Annu. Rev. Polit. Sci.}, 11:\penalty0 387--412, 2008.

\bibitem[van Bokkem et~al.(2019)van Bokkem, Hageman, Koning, Nguyen, and
  Zarin]{Bokkem2019}
Dirk van Bokkem, Rico Hageman, Gijs Koning, Luat Nguyen, and Naqib Zarin.
\newblock Self-sovereign identity solutions: The necessity of blockchain
  technology.
\newblock \emph{arXiv preprint arXiv:1904.12816}, 2019.

\bibitem[Victor and Zickau(2018)]{Victor2018}
Friedhelm Victor and Sebastian Zickau.
\newblock Geofences on the blockchain: Enabling decentralized location-based
  services.
\newblock In \emph{2018 IEEE International Conference on Data Mining Workshops
  (ICDMW)}, pages 97--104. IEEE, 2018.

\bibitem[Vosoughi et~al.(2018)Vosoughi, Roy, and Aral]{Vosoughi2018}
Soroush Vosoughi, Deb Roy, and Sinan Aral.
\newblock The spread of true and false news online.
\newblock \emph{Science}, 359\penalty0 (6380):\penalty0 1146--1151, 2018.

\bibitem[Wan et~al.(2010)Wan, Torelli, and Chiu]{Wan2010}
Ching Wan, Carlos~J Torelli, and Chi-yue Chiu.
\newblock Intersubjective consensus and the maintenance of normative shared
  reality.
\newblock \emph{Social Cognition}, 28\penalty0 (3):\penalty0 422--446, 2010.

\bibitem[Weatherford(1992)]{Weatherford1992}
M~Stephen Weatherford.
\newblock Measuring political legitimacy.
\newblock \emph{American political science review}, 86\penalty0 (1):\penalty0
  149--166, 1992.

\bibitem[Wolberger and Fedyukovych(2018)]{Wolberger2018}
L~Wolberger and V~Fedyukovych.
\newblock {Zero Knowledge Proof of Location--Platin ZK Yellow Paper}.
\newblock
  https://platin.io/assets/yellowpaper/Platin\_Yellow\_Paper\_2018-11-18.pdf,
  2018.
\newblock (last accessed: April 2020).

\bibitem[Yasaweerasinghelage et~al.(2017)Yasaweerasinghelage, Staples, and
  Weber]{Yasaweerasinghelage2017}
Rajitha Yasaweerasinghelage, Mark Staples, and Ingo Weber.
\newblock Predicting latency of blockchain-based systems using architectural
  modelling and simulation.
\newblock In \emph{2017 IEEE International Conference on Software Architecture
  (ICSA)}, pages 253--256. IEEE, 2017.

\bibitem[Yucel and Bulut(2018)]{Yucel2018}
Fatih Yucel and Eyuphan Bulut.
\newblock Clustered crowd gps for privacy valuing active localization.
\newblock \emph{IEEE Access}, 6:\penalty0 23213--23221, 2018.

\bibitem[Zhu and Cao(2013)]{Zhu2013}
Zhichao Zhu and Guohong Cao.
\newblock Toward privacy preserving and collusion resistance in a location
  proof updating system.
\newblock \emph{IEEE Transactions on Mobile Computing}, 12\penalty0
  (1):\penalty0 51--64, 2013.

\bibitem[Ziegler et~al.(2019)Ziegler, Gro$\beta$mann, and Krieger]{Ziegler2019}
Michael~Herbert Ziegler, Marcel Gro$\beta$mann, and Udo~R Krieger.
\newblock Integration of fog computing and blockchain technology using the
  plasma framework.
\newblock In \emph{2019 IEEE International Conference on Blockchain and
  Cryptocurrency (ICBC)}, pages 120--123. IEEE, 2019.

\bibitem[Zolotov et~al.(2018)Zolotov, Oliveira, and Casteleyn]{Zolotov2018}
Mijail~Naranjo Zolotov, Tiago Oliveira, and Sven Casteleyn.
\newblock E-participation adoption models research in the last 17 years: A
  weight and meta-analytical review.
\newblock \emph{Computers in Human Behavior}, 81:\penalty0 350--365, 2018.

\end{thebibliography}


\newpage 

\bio{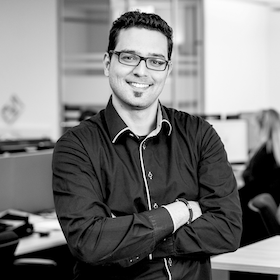}
Dr. Evangelos Pournaras is an Associate Professor at Distributed Systems and Services group, School of Computing, University of Leeds, UK. He is also currently a research associate at UCL Center of Blockchain Technologies and research fellow in blockchain industry. He has more than 5 years experience as senior scientist and postdoctoral researcher at ETH Zurich in Switzerland after having completed his PhD studies in 2013 at Delft University of Technology and VU University Amsterdam in the Netherlands. Evangelos has also been a visiting researcher at EPFL in Switzerland and has industry experience at IBM T.J. Watson Research Center in the USA. Since 2007, he holds a MSc with distinction in Internet Computing from University of Surrey, UK and since 2006 a BSc on Technology Education and Digital Systems from University of Piraeus, Greece. Evangelos has won the Augmented Democracy Prize, the 1st prize at ETH Policy Challenge as well as 4 paper awards and honors. He has published more than 50 peer-reviewed papers in high impact journals and conferences and he is the founder of the EPOS, DIAS, SFINA and Smart Agora projects featured at decentralized-systems.org. He has raised significant funding and has been actively involved in EU projects such as ASSET, SoBigData and FuturICT 2.0. He has supervised several PhD and MSc thesis projects, while he designed courses in the area of data science and multi-agent systems that adopt a novel pedagogical and learning approach. Evangelos' research interest focus on distributed and intelligent social computing systems with expertise in the inter-disciplinary application domains of Smart Cities and Smart Grids.
\endbio

\end{document}